\documentclass[conference]{IEEEtran}
\usepackage{blindtext}
\usepackage{multicol}
\usepackage{cite}
\usepackage{amsmath,amssymb,amsfonts}
\usepackage{algorithmic}
\usepackage{graphicx}
\usepackage{subcaption}
\usepackage{sidecap}
\usepackage{placeins}
\usepackage{textcomp}
\usepackage{notoccite}
\usepackage[utf8]{inputenc} % usually not needed (loaded by default)
\usepackage[T1]{fontenc}


\begin{document}

\author{
	\IEEEauthorblockN{Matthew Skrzypczyk\IEEEauthorrefmark{1} and Stephanie Wehner\IEEEauthorrefmark{1}}
	\IEEEauthorblockA{\IEEEauthorrefmark{1}QuTech and Kavli Institute of Nanonscience, 2628 CJ Delft, Netherlands\textsuperscript{\textsection}}
}

\title{An Architecture for Meeting Quality-of-Service Requirements in Multi-User Quantum Networks}

\maketitle

\begingroup\renewcommand\thefootnote{}
\footnotetext{Corresponding authors: Matthew Skrzypczyk (email: mdskrzypczyk@tuta.io) and Stephanie Wehner (email: S.D.C.Wehner@tudelft.nl)}

\begingroup\renewcommand\thefootnote{\textsection}
\footnotetext{We acknowledge financial support from the EU Flagship on Quantum Technologies through the project Quantum Internet Alliance (EU Horizon 2020, grant agreement no.820445).}

\begin{abstract}
Quantum communication can enhance internet technology by enabling novel applications that are provably impossible classically. The successful execution of such applications relies on the generation of quantum entanglement between different users of the network which meets stringent performance requirements. Alongside traditional metrics such as throughput and jitter, one must ensure the generated entanglement is of sufficiently high quality. Meeting such performance requirements demands a careful orchestration of many devices in the network, giving rise to a fundamentally new scheduling problem.  Furthermore, technological limitations of near-term quantum devices impose significant constraints on scheduling methods hoping to meet performance requirements.
In this work, we propose the first end-to-end design of a centralized quantum network with multiple users that orchestrates the delivery of entanglement which meets quality-of-service (QoS) requirements of applications. We achieve this by using a centrally constructed schedule that manages usage of devices and ensures the coordinated execution of different quantum operations throughout the network.  We use periodic task scheduling and resource-constrained project scheduling techniques, including a novel heuristic, to construct the schedules. Our simulations of four small networks using hardware-validated network parameters, and of a real-world fiber topology using futuristic parameters, illustrate trade-offs between traditional and quantum performance metrics.
\end{abstract}

\section{Introduction}
\label{sec:introduction}
Recent progress in developing networked quantum devices (see e.g. ~\cite{hofmann2012heralded, reiserer2016robust, humphreys2018deterministic, krutyanskiy2019light, Pan50kms, PanSatellite}) motivates the emerging field of quantum network architecture.
% in order to scale such networks. 
Quantum networks promise to significantly enhance internet technology by enabling new applications that are impossible to achieve using classical (non-quantum) communication~\cite{kimble2008quantum,wehner2018quantum}.
Key to enabling quantum applications is the creation of end-to-end entanglement between two nodes in the network. Entanglement is a special property of two quantum bits (qubits) held by two nodes in the quantum network. As such, one might think of entanglement as a form of virtual or - \emph{entangled - link} between the two qubits~\cite{SchouteRouting}.
% to refer to two entangled qubits at different nodes in the network. 

Application performance in quantum networks depends on several dimensions of network service. On one hand, there are traditional performance metrics such as the \emph{throughput} of entanglement delivery as well as \emph{jitter} (variance in inter-delivery times) for more complex applications~\cite{dahlberg2019link}. On the other hand, there is a genuinely quantum performance metric, namely the quality, or \emph{fidelity}, of the entanglement delivered to users~\cite{dahlberg2019link}.
Meeting application requirements and maximizing network utility motivates the design of quantum network architectures that support quality-of-service (QoS) guarantees on the distributed entanglement.

Entanglement, or \emph{entangled links}, may be established between quantum network devices that are directly connected via a physical medium such as optical fiber (see e.g.~\cite{humphreys2018deterministic, krutyanskiy2019light, Pan50kms}), or free-space communication (see e.g.~\cite{PanSatellite}).
We refer to two such devices as \emph{connected}. 
In multi-hop quantum networks, where not all devices are connected, entanglement distribution can be accomplished with the help of intermediary nodes using a procedure known as \emph{entanglement swapping} (see Section~\ref{sec:quantumDevices}). Such intermediary quantum devices are often referred to as a \emph{quantum repeater}. 
In general, quantum repeater protocols that establish entanglement over long distances can be formed by combining several types of operations in addition to entanglement swapping (see Section~\ref{sec:quantumDevices}). The fidelity requirement on entanglement distribution is satisfied by the exact combination of these operations, as allowed by the underlying quantum hardware. Throughput requirements are met by executing quantum repeater protocols frequently enough to distribute entanglement at the desired rate while jitter requirements are met by regulating the inter-delivery times of entanglement from the quantum repeater protocols.

Even if only two users in the network wish to communicate, the successful execution of a quantum repeater protocol requires the coordinated execution of different operations at intermediary nodes in the network (see Section~\ref{sec:quantumDevices}). If many users wish to generate entanglement simultaneously, we also require coordination between the actions of two disjoint repeater protocols at the level of their component operations. This gives rise to a novel scheduling problem that is fundamental to the design of quantum networks. 

What is more, near-term technological limitations impose very strict demands on any coordination mechanism hoping to meet QoS requirements. Specifically, near-term quantum hardware (sometimes also referred to as noisy intermediate-scale quantum devices, NISQ~\cite{preskill2018quantum}) offers limited memory lifetimes (at most seconds \cite{abobeih2018one,bradley2019ten}), which means that entangled links cannot be stored for a long time. Furthermore, a limited storage space means that the number of entangled links that can be stored simultaneously is small (only recently $2$~\cite{upcomingGHZ}). These limitations impose both real-time and resource constraints on the creation of entangled links.
%As a result, NISQ devices severely limit network performance, and will neccessitate the use of strict coordination mechanisms to meet the QoS requirements of network applications. 

Here, we propose a novel time-division multiple access (TDMA) network architecture for quantum networks that supports QoS requirements of entanglement generation for applications. Our centralized architecture achieves this by encoding quantum repeater protocols into schedules that are distributed across the network. Fixed-duration time slots in the schedule encode the different operations of quantum repeater protocols.
The encoded protocols are selected to succeed with high probability and meet fidelity requirements, while the schedule is constructed such that the frequency of entanglement delivery meets throughput and jitter requirements (see Section \ref{sec:sched_med} for details). To this end, we introduce the novel problem of constructing schedules of quantum repeater protocols and provide several methods, including a new heuristic, for solving the resulting scheduling problem. We benchmark our new heuristic against existing heuristics adaptable to this setting on several small near-term quantum network topologies as well as a futuristic network based on a real-world fiber topology in the Netherlands and show that comparable performance can be obtained while reducing runtime complexity. We additionally find that the choice of scheduling heuristic can be used to trade off higher network throughput for lower jitter. 
We emphasize that the use of a centralized architecture to coordinate entanglement distribution has no effect on quantum security applications (e.g. quantum key distribution~\cite{bennett1984proceedings,ekert1991quantum}) as the network in its entirety is treated as  untrusted in their security analysis.

We design our architecture to fit into an existing quantum network stack, and build upon previously proposed quantum network protocols~\cite{dahlberg2019link,kozlowski2020designing}. In particular, our architecture can work seemlessly in conjunction with the link layer protocol proposed in~\cite{dahlberg2019link} where it is used to replace the agreement functionality provided by the distributed queue protocol~\cite{dahlberg2019link}, thus allowing scalability to larger networks. This is showcased by our recent realization of the link layer protocol on quantum hardware \cite{PompiliExperimental}, which uses a (highly simplified) form of the TDMA architecture proposed here. Our architecture has the following defining properties:
(1) encoded repeater protocols are connection-oriented and can be tailored per application, (2)
the centrally constructed schedule provides contention-free usage of network devices, (3)
dynamic update of the schedule allows the system to accompany network demands at runtime.  Our contributions can be summarized as follows:

\begin{itemize}
	\item We introduce the first centralized quantum network architecture that supports QoS requirements for multiple concurrent users (Section \ref{sec:arch}),
	\item we develop a repeater protocol model that allows calculation of success probability and worst-case end-to-end fidelity (Section \ref{sec:rep_protocol_model}),
	\item we introduce two scheduling methods based on periodic task scheduling and RCPSP for constructing network-wide schedules of repeater protocols that satisfy QoS requirements on entanglement delivery (Section \ref{sec:sched}),
	\item we introduce a novel heuristic for the RCPSP-based scheduling method which achieves comparable performance while reducing runtime complexity from $O(N^2 S^2 |K| \log(NS))$ to $O(N^2 |K| \log(N))$ (Section \ref{sec:sched_rcpsp}),
	\item we evaluate our scheduling methods and benchmark our new heuristic against existing heuristics for our proposed methods (Section \ref{sec:eval}).
\end{itemize}

% The remainder of this paper is structured as follows.  Section \ref{sec:relwork} will briefly summarize related work in the domains of quantum network architectures as well as TDMA network architectures. Section \ref{sec:des_con} will discuss our design considerations of quantum networks. We present our proposed architecture in Section \ref{sec:arch} and detail the problem of schedule construction in Section \ref{sec:sched}. We evaluate and compare the performance of several scheduling heuristics in Section \ref{sec:eval} and conclude the paper in Section \ref{sec:conc}.

\section{Related Work}
\label{sec:relwork}
Several functional allocations of quantum network stacks have been proposed in \cite{aparicio2011protocol,van2014quantum,van2008system,van2013designing} that formulate layers based on specific protocols such as entanglement distillation. In contrast, Dahlberg et al. take a different approach in \cite{dahlberg2019link} and focus on the type of service each layer should provide. The authors complement their functional allocation with physical and link layer protocols that take practical considerations, such as hardware imperfections and communication overhead, into account. Kozlowski et al. \cite{kozlowski2020designing} build upon this work and design a network layer protocol suitable for the network stack model of \cite{dahlberg2019link}. In \cite{matsuo2019quantum}, Matsuo et al. present and simulate a RuleSet-based quantum link bootstrapping protocol that may be used to install rules to provide flexibility when establishing connections in quantum networks.

Quantum network architectures that break down entanglement distribution into discrete time steps have been studied in \cite{das2018robust, pant2019routing,shi2020concurrent}. In contrast to our paper, these works do not describe the network infrastructure that realizes their architecture nor do they detail any scheduling strategies for network coordination. Furthermore, they depend on the production of high quality entanglement between connected devices in order to meet high fidelity requirements between distant nodes in the network and as a result are not suitable for near-term networks that we consider here.  

Cicconiti et al. consider a time-slotted, centralized quantum network architecture and the problem of request scheduling, though their approach differs from ours in that it does not aim to satisfy QoS requirements of applications at end nodes.  Additionally, the architecture presented actively communicates with network nodes to decide which available entangled links should be used to service network demands.  In contrast, we fix the path and set of entangled links used to service the throughput, jitter, and fidelity requirements of each pair of users in the network.  
% REPHRASE THIS
% These works deliver entanglement to end nodes by first attempting entanglement establishment between connected pairs of devices and then performing entanglement swapping over a path of entangled links that were successfully established connecting the end nodes. While these architectures can establish entanglement between end nodes, they are not suitable for meeting high fidelity requirements when entangled links between connected devices are established with low fidelity. Meeting fidelity and rate requirements in this setting necessitates the use of low-latency quantum repeater protocols that utilize entanglement distillation to increase fidelity. Our proposed architecture allows this by predetermining the path and the appropriate quantum repeater protocol that meets desired fidelity and rate requirements.
In \cite{aparicio2011multiplexing}, Aparicio et al. evaluate the usage of time-division multiplexing (TDM) in comparison to other multiplexing strategies. Their TDM scheme differs from ours in that it assigns end-to-end flows to time slots without detailing the quantum repeater protocol operations to execute. As a result, their scheme does not coordinate the operation of connected devices that are limited to establishing one entangled link with another connected device at a time. Vardoyan et al. study the performance of a quantum network switch within a star topology in \cite{vardoyan2019capacity} and find that correctly configured scheduling policies can outperform TDM for smaller star topologies, but that the relative improvement reduces as the number of users grows.

Dynamic TDMA protocols have been studied in the context of many network technologies: ad-hoc networks \cite{kamruzzaman2010dynamic}, wireless ATM networks \cite{frigon2001dynamic}, wireless powered communication networks \cite{kang2014optimal}, and satellite communication~\cite{petraki2007dynamic}. An extensive amount of literature on TDMA schemes ranging from centralized to distributed models \cite{hossain2001centralized, salonidis2005distributed} exists. In contrast, our TDMA architecture for quantum networks multiplexes the execution of operations for quantum repeater protocols, of which there may be many on a single node just to establish a single end-to-end entangled link, rather than multiplexing access to classical channels. 
%Quantum networks augment the capabilities of traditional networks and thus classical communication is assumed to exist between network nodes.

Construction of TDMA schedules is a long-studied problem in several types of networks. Traditionally, TDMA schedules are constructed in order to enable data transmission between nodes in networks over shared communication mediums. Studied methods include formulating schedule construction as a graph coloring problem \cite{ramanathan1997unified} and using heuristics \cite{kang2013graph} or branch-and-bound search \cite{bertossi1987time} techniques to compute a solution. These existing formulations are not suitable for NISQ device quantum networks as end-to-end transmission in classical networks may be achieved by preventing collisions on common communication channels whereas establishing end-to-end entanglement with sufficient fidelity requires coordinating operations along the entire path through the network.
% is inherently different from data transmission in classical networks.

A related problem to schedule construction is that of scheduling task graphs to parallel processor systems \cite{ahmad1996analysis} where parallel programs represented by directed acyclic graphs are allocated to a set of homogeneous processors. This also differs from our scheduling problem as operations in repeater protocols must be allocated to specific network nodes in order to meet QoS requirements.

\FloatBarrier
\begin{figure*}[h!]
	\centering
	\begin{subfigure}{0.4\linewidth}
		\caption{}
		\includegraphics[width=\textwidth]{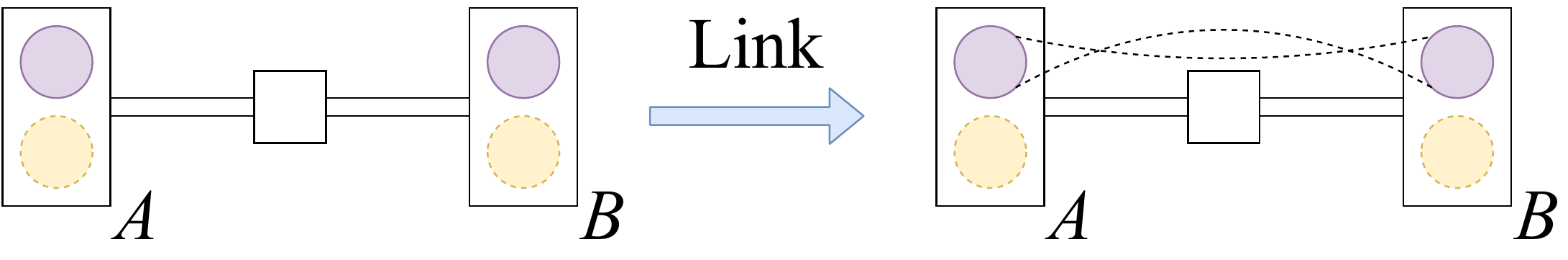}
		\label{fig:ent_gen}
	\end{subfigure}%
	\hspace{0.1\linewidth}%
	\begin{subfigure}{0.4\linewidth}
		\caption{}
		\includegraphics[width=\textwidth]{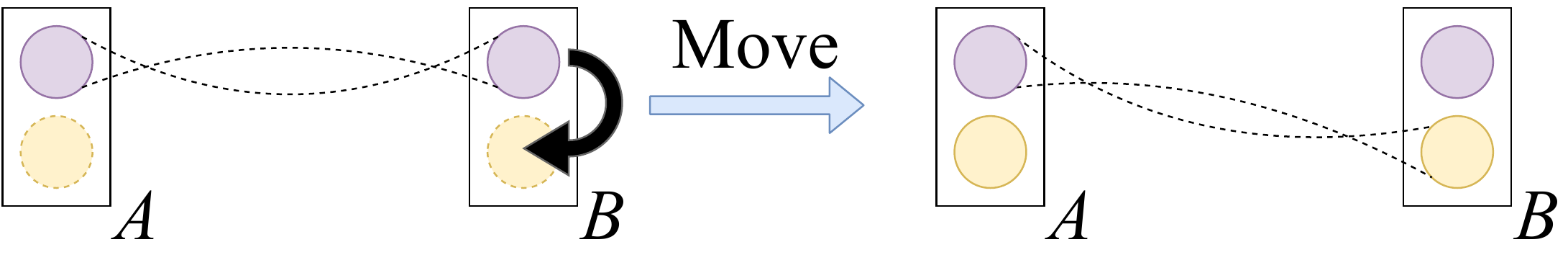}
		\label{fig:move}
	\end{subfigure}
	\begin{subfigure}{0.4\linewidth}
		\caption{}
		\includegraphics[width=\textwidth]{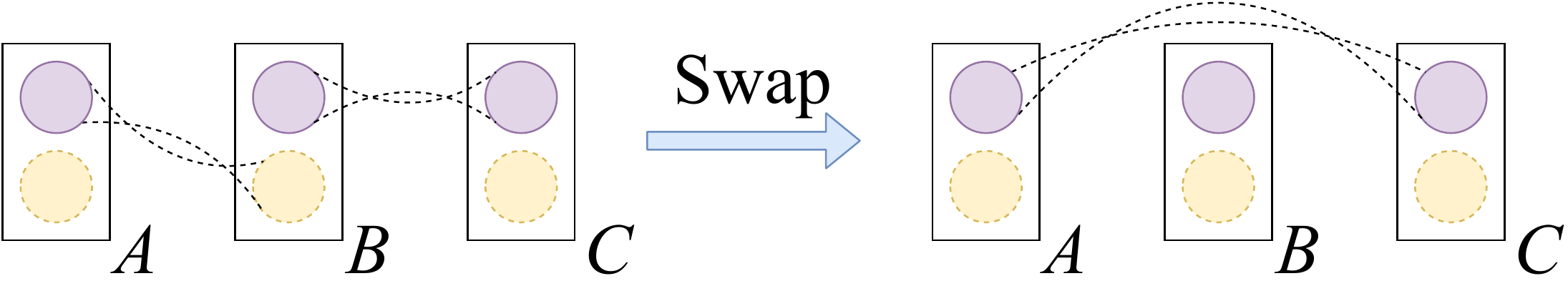}
		\label{fig:swap}
	\end{subfigure}%
	\hspace{0.1\linewidth}%
	\begin{subfigure}{0.4\linewidth}
		\caption{}
		\includegraphics[width=\textwidth]{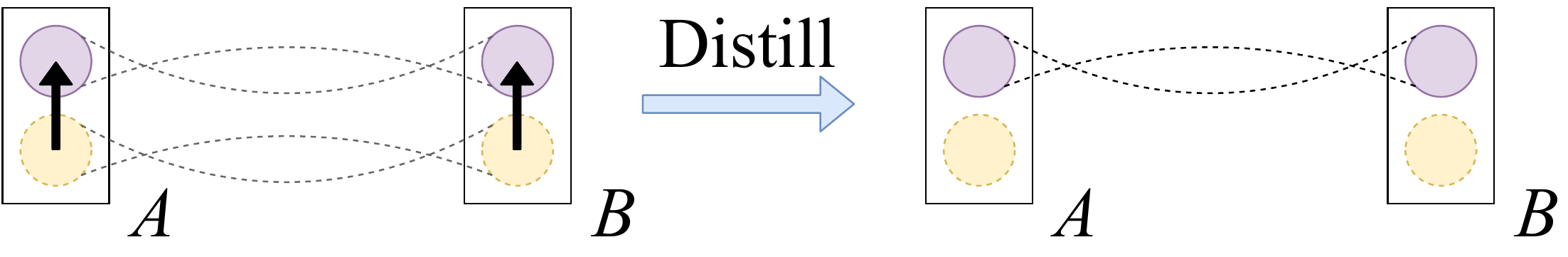}
		\label{fig:distill}
	\end{subfigure}
	\caption{Basic quantum repeater operations to produce long-distance entanglement. a) Generation of an elementary entangled link between two devices connected by a physical medium.
		b) Memory operation: Certain quantum devices may move the qubit to a different location in memory in order to generate further entangled links. c) Entanglement swapping can be used to produce an entangled link between two unconnected quantum devices. d) Entanglement distillation can be used to produce one (or more) entangled links with a higher fidelity (quality) from two (or more) links of lower quality.}
	\label{fig:basics}
\end{figure*}

\begin{SCfigure*}
	\centering
	\includegraphics[width=1.1\linewidth]{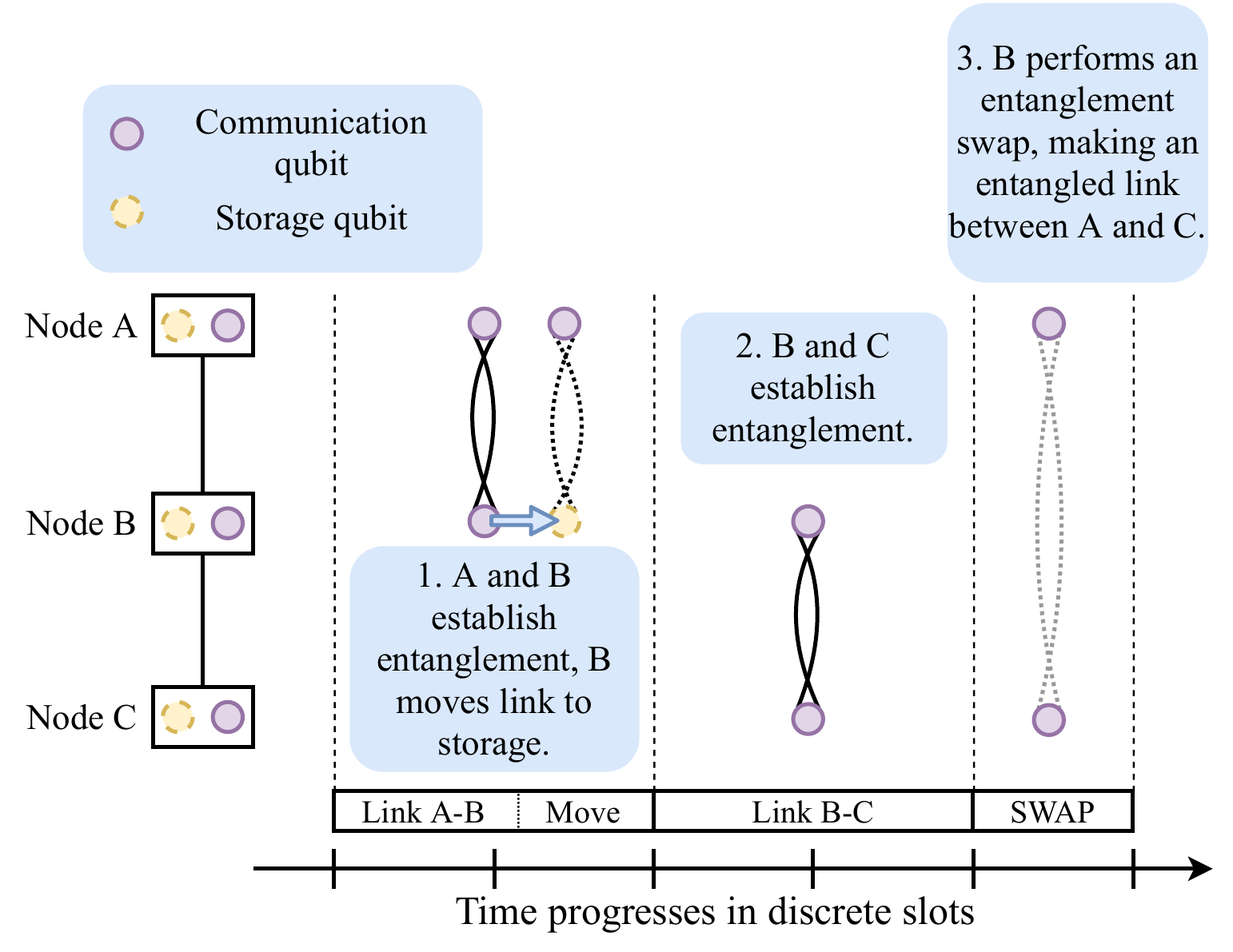}
	\caption{Temporal visualization of a quantum repeater protocol establishing one entangled link between two nodes $A$ and $C$ via one \emph{intermediary node} $B$ (e.g. a quantum repeater). All nodes perform operations according to a schedule that allocates operations to specific time slots.
		The first operation produces an elementary entangled link between $A$ and $B$, followed by a move operation to the storage qubit. Operations 
		may be assigned into time slots of sufficient length to produce entanglement with a heralding protocol (see Section \ref{sec:sched_constr} for a discussion). Due to limited parallelism at device $B$, the elementary entangled link $B-C$ can only be produced in the subsequent time period. Once the time slots of producing both elementary links w.h.p. and executing the move have elapsed, an entanglement swapping operation is executed, consuming the original entanglement to produce one entangled link $A-C$.
	}
	\label{fig:example_rep_protocol}
\end{SCfigure*}

\section{Design Considerations}
\label{sec:des_con}
Quantum networks pose several new challenges to address when designing the network's architecture.  Design considerations for producing entangled links between connected nodes can be found in~\cite{dahlberg2019link}. We hence focus mainly on the challenges posed by long-distance entanglement generation as relevant to the design of our TDMA architecture.

\subsection{Devices and Establishing Entanglement}\label{sec:quantumDevices}
A quantum network consists of \emph{end nodes}~\cite{wehner2018quantum} that are connected to the network in order to run specific applications. 
%These may be simple photonic devices capable of measuring only a single qubit at a time, or full blown quantum processors capable of universal quantum computation. 
In addition, a quantum network may include nodes, known as \emph{repeater nodes}, that facilitate the generation of entanglement between two unconnected nodes.  End nodes can also act as repeater nodes, but QoS demands only originate from applications at end nodes.
Here, we focus primarily on quantum nodes (end nodes or repeaters) known as processing nodes. 
A processing node is a few-qubit quantum computer with an optical interface, of which there have been implementations in nitrogen vacancy (NV) centers in diamond~\cite{hensen2015loophole}, ion traps~\cite{moehring2007entanglement} and neutral atoms~\cite{hofmann2012heralded}.
We emphasize however that our work can also be adapted to other platforms, including systems based on atomic ensembles~\cite{chou2005measurement}. 

To produce long-distance entanglement, near-term quantum repeater protocols~\cite{LiangJiangGenerations} may employ a variety of operations (Fig.~\ref{fig:basics}) which need to be scheduled in a coordinated manner (see e.g. Fig.~\ref{fig:example_rep_protocol}).
Two directly connected nodes can establish an \emph{elementary entangled link} between them (Fig.~\ref{fig:basics}a).
Entanglement generation at the physical layer is probabilistic, and will often require several attempts before an entangled link is created. 
Here, we focus on using a robust link layer protocol~\cite{dahlberg2019link} based around a physical layer entanglement generation scheme that has a \emph{heralding signal} confirming the success/failure.
This link layer protocol allows for a deliberate trade-off between the fidelity and the throughput of generating elementary links, depending on the capabilities of the underlying hardware.
%We refer to these schemes as \emph{heralded entanglement generation} schemes. 
Coordinating establishment of a single entangled link between connected nodes already requires timing synchronization and agreement~\cite{dahlberg2019link} (up to ns precision using e.g. White Rabbit~\cite{serrano2013white}). 
%Chains of automated nodes \cite{dahlberg2019link} that produce entangled links with fixed fidelity between its endpoints are treated as elementary entangled links.

In multi-hop quantum networks, entanglement distribution can be accomplished with the help of intermediary nodes using an operation known as \emph{entanglement swapping} (see Fig. \ref{fig:basics}c): 
two nodes that share no physical connection (A and C) first produce an elementary entangled link with the intermediary node (B) via their shared physical medium. The fidelity requirement for such elementary links can be computed from the end-to-end QoS requirements. When both entangled links are ready, a measurement (the swapping operation) can be performed on both qubits at node B, which produces end-to-end entanglement between two unconnected devices A and C. This measurement consumes the original entangled links. Since entanglement generation is a probabilistic process, forming long distance entangled links efficiently asks for B to have a quantum memory in which the qubit of one entangled link (say the link A-B) can be stored until the second link (say B-C) is ready. Entanglement swapping operations can either be probabilistic, or deterministic depending on the underlying physical implementation. While our work can also be extended to systems in which swapping operations are probabilistic, we here focus on the case of deterministic swaps as allowed by processing nodes built from eg. NV centers in diamond, ion traps, or neutral atoms.
When producing entanglement between two end nodes that are separated by one (or more) intermediary node(s), failure to achieve QoS requirements in any one entangled link in the chain leads to a failure to achieve overall end-to-end QoS.
It is important that the swapping operation is applied to the correct entangled links, requiring the careful allocation of qubits to protocol operations across the network. Otherwise, we may establish a link between incorrect pairs of users (or none at all) resulting in application failure. 

To meet fidelity requirements, quantum repeater protocols may also employ entanglement distillation operations~\cite{duerSurvey} which turn multiple low-fidelity links into a fewer number of higher quality links. Entanglement distillation has been demonstrated using NV centers in diamond~\cite{kalb2017entanglement}. This operation is in general probabilistic, but a heralding signal is produced to indicate success or failure~\cite{kalb2017entanglement}. For all such repeater operations, the nodes need to exchange classical information, and we hence take a classical network supporting entanglement generation as a given.

Several technological limitations impose stringent demands on any coordination mechanism used to produce end-to-end entanglement. 
First, due to limited lifetimes of quantum memories, the fidelity of an entangled link decreases exponentially with the storage time (at most seconds~\cite{abobeih2018one}). 
Repeater nodes that lack the ability to store entanglement, or have very short storage times therefore must establish the needed links close in time. Any schedule that does not ensure that the links are produced close in time (i.e. missing deadlines) for any single hop in the chain connecting the two end nodes will hence lead to a failure in end-to-end entanglement generation with the desired QoS requirements. 

Second, NISQ devices can only store a limited amount of quantum information at a time. This limits the number of entangled links that a node can hold simultaneously (demonstrated $2$~\cite{upcomingGHZ}), posing additional resource allocation challenges. Processing nodes typically have different types of qubits~\cite{dahlberg2019link}: communication qubits with an optical interface for entanglement generation with connected nodes, as well as storage qubits which can solely be used for storing and manipulating qubits in memory. As such, quantum repeater protocols may necessitate a move operation in memory (Fig.~\ref{fig:basics}b).

Third, several network device platforms for processing end nodes (see e.g.\cite{humphreys2018deterministic}) are limited to either processing qubits (e.g. swapping or distilling operations), or establishing entanglement exclusively at any time but not both. 
%This means that such a node can at any one time either establish entanglement with a neighboring node, or perform other operations such as entanglement swapping or distillation. 
This imposes additional timing constraints in any schedule.

% Several experimental realizations of network-enabled quantum devices exist \cite{hensen2015loophole,moehring2007entanglement,hofmann2012heralded,delteil2016generation,valivarthi2016quantum} and current state-of-the-art devices fall under the classification of NISQ \cite{preskill2018quantum}. These devices are characterized by imperfect qubit control, limited qubit counts, and limited qubit lifetimes. In terms of quantum repeater protocol execution, these devices introduce a number of challenges in long-distance entanglement delivery.

%First, entanglement swapping requires simultaneous availability of two entangled links. Repeater nodes that lack the ability to store entanglement or have very short storage times must establish the needed links close in time (at present within $0.542$ ms \cite{jobez2016towards} to $0.53$ s \cite{holzapfel2020optical}). In this work, we focus on entanglement establishment schemes based around a \emph{heralding signal} that confirms the success/failure of entangling qubits between two connected quantum network devices (several schemes exist, see \cite{dahlberg2019link} for an example implementation). We refer to these schemes as \emph{heralded entanglement generation} schemes. These entanglement establishment schemes are probabilistic and will often require several attempts before an entangled link is created. Coordinating establishment of a single entangled link is already a non-trivial process \cite{dahlberg2019link} and extending this coordination to multiple entangled links over several network devices complicates the problem further.

\subsection{Application Requirements}
\label{sec:app_requirements}
Much like applications in traditional networks, applications in quantum networks may observe different traffic patterns and quality-of-service (QoS) requirements in order to execute correctly. Applications of the \emph{Measure Directly} ($MD$)  use case~\cite{dahlberg2019link} produce many end-to-end entangled links, but do not require them to be stored nor produced at the same time. In this case, throughput may be a strictly required QoS metric while jitter has no impact on application performance. In contrast, some applications of the \emph{Create and Keep} ($CK$) use case \cite{dahlberg2019link} may require storing multiple entangled links at the same time. Since memory lifetimes are short, applications of $CK$ use cases may observe stricter jitter requirements in order to ensure that sufficiently many entangled links can be produced within the same time window. In all use cases, requirements on entanglement fidelity vary depending on the error tolerance of applications, providing flexibility in the choice of quantum repeater protocols for delivering entanglement. Designing quantum networks that meet varying levels of QoS requirements thus increases the number of supported applications and consequently its utility.

\section{TDMA Architecture Design}
\label{sec:arch}
% Designing a network architecture that is capable of satisfying QoS requirements given the previously mentioned design considerations is a non-trivial problem. Here, we present our architecture and discuss how we address our design considerations.

% In this section we present our proposed TDMA architecture for quantum networks and show how the design considerations outlined in Section \ref{sec:des_con} are addressed.  We will first explain how we represent QoS requirements as network demands that applications communicate to the network and then detail how network nodes use a TDMA schedule to coordinate the execution of quantum repeater protocols. We then describe the network infrastructure that realizes our architecture before concluding with a summary of failure-handling mechanisms.

\begin{figure*}[htb!]
	\centering
	\includegraphics[width=\textwidth]{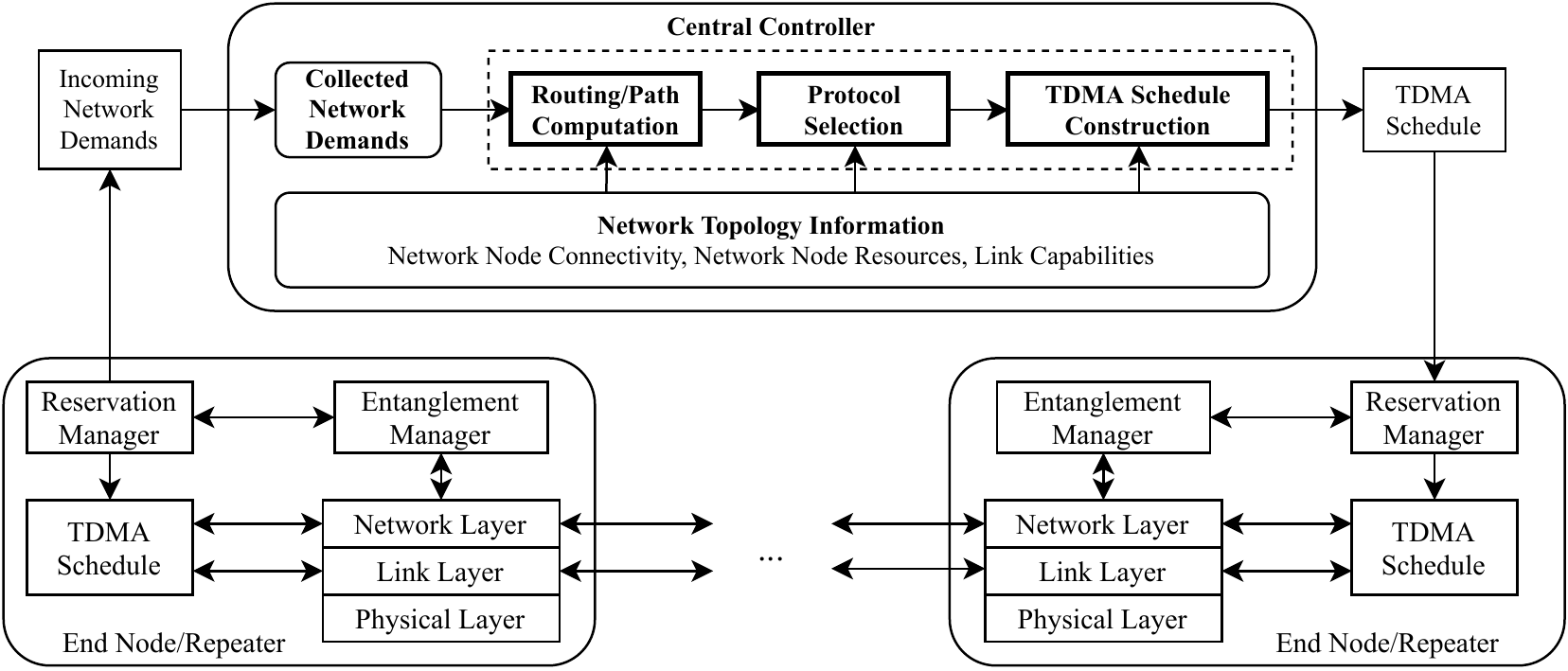}
	\caption{Interaction diagram of software components of end nodes and repeater nodes in the quantum network. Ellipses denote additional network nodes with the same software while arrows denote communication between software components. A reservation manager acts as an interface between applications and the central controller for expressing network demands while the entanglement manager tracks delivered entanglement and provides it to requesting applications. Network demands are forwarded to a central controller where they are subject to admission control and used to produce a new network schedule. The reservation manager installs these schedules for use by the local quantum network stack.}
	\label{fig:network_stack}
\end{figure*}

\subsection{Centralized Control}
Due to the design considerations posed by near-term quantum devices, we opt for a centralized architecture in which a central controller is responsible for setting a network-wide schedule for end-to-end entanglement generation based on demands communicated by the end nodes (see below). This allows us to mitigate limited memory lifetimes imposing strict deadlines on the schedule of repeater protocol operations while maximizing network usage for many users. In addition, a network-wide schedule provides a ready means of tracking which entangled links are used in swapping operations, ensuring entanglement is created between the correct nodes. This removes the need for real-time discussion among network nodes to coordinate quantum operations and reduces the complexity of implementation.

Before any quantum communication takes places, the nodes engage in a discussion with the central controller who acts as a repository of all information required to schedule repeater protocols. The controller holds information such as the network topology and link capabilities, i.e., the available choices of fidelity, throughput and latency at which an elementary link can be produced for each pair of connected nodes~\cite{dahlberg2019link}. Such topology information may be acquired using link bootstrapping protocols (see e.g.~\cite{dahlberg2019link,matsuo2019quantum}) for characterizing QoS capabilities between connected nodes. What is more, the information includes hardware capabilities of the individual nodes themselves, i.e. their available communication and storage qubits, the quality and speed of their operations affecting end-to-end QoS capabilities, as well as their availability for producing entanglement in given time slots. This information should be updated intermittently to keep the central controller up-to-date as near-term NISQ devices may require intermittent calibration and the capabilities of links may drift with time. The central controller has no quantum capabilities and does not participate in any quantum repeater protocol, though it requires timing synchronization with network nodes for coordinating changes to the network schedule. To prevent disruptions in network service, such a central controller should be realized using a fault-tolerant distributed computing architecture \cite{jalote1994fault}.

\subsection{Starting Communication}
Before entanglement generation for a specific application on end nodes $A$ and $B$ commences, $A$ and $B$ form a classical connection to
agree on QoS requirements that entanglement generation should obey. 
These demands for entanglement are then communicated to the controller along with a maximal time that the end nodes are willing to wait before entanglement production starts. End nodes may also request the central controller to exclude slots so as to allow time for processing entanglement between scheduled operations. The controller then produces a schedule that captures QoS requirements. If demands exceed network capabilities or cannot be fulfilled within the desired time, the controller rejects the new demands. Once entanglement generation starts, the central controller will schedule the requested demand continuously until the end nodes ask to stop entanglement production.

Integration into the quantum network stack of~\cite{dahlberg2019link} can be achieved as in Fig.~\ref{fig:network_stack}:
User applications at the end nodes $A$ and $B$ communicate their requirements using a \emph{reservation manager} that acts as an interface between the end node and the central controller (Fig. \ref{fig:network_stack}). This reservation manager provides a service interface akin to the link layer interface in~\cite{dahlberg2019link}, allowing applications to specify requirements such as fidelity $F$, the desired number of entangled pairs $N$ (or throughput $R$), and constraints on jitter $J$. If needed, the reservation manager translates $N$ into a rate $R$.
This manager is responsible for submitting all application specific network demands $(A, B, F, R, J)$ to the central controller, and for installing schedules for local use. The reservation manager may additionally supplement the demands with a specification of time needed to process entanglement, allowing slots to be excluded from the schedule so that end nodes can process entanglement before subsequent repeater protocol operations. If network demands are accepted by the central controller, an \emph{entanglement manager} serves the created entangled links to the applications in accordance with their requests, 
%That is, requests to create entanglement~\cite{dahlberg2019link} as part of an application are first passed to the entanglement manager who fulfills them 
whenever entanglement according to the central controller's schedule becomes ready. The network layer~\cite{kozlowski2020designing} is responsible for executing both swapping and distillation operations and must communicate the associated control information, including e.g. measurement outcomes (see Section \ref{sec:quantumDevices}), from these operations to other network nodes. The link layer~\cite{dahlberg2019link} is used to produce elementary entangled links with connected nodes, where the distributed queue of~\cite{dahlberg2019link} is replaced with the central network schedule. When the desired number of entangled links has been produced (or an application no longer desires entanglement), end nodes immediately cease to participate in the operations scheduled to serve the specific application, and the reservation manager updates the central controller.

% In this paper, we consider QoS requirements on the throughput and fidelity of entanglement delivery and express these QoS requirements as a network demand. A network demand $D$ is represented by a tuple $(src,dst,F_{min},R_{min})$ denoting the source and destination end nodes to establish entanglement between as well as the minimum fidelity and rate of entanglement delivery. End nodes formulate network demands based on their application requirements and submit these demands to a centralized controller that creates and distributes TDMA schedules. We use the terminology of a \emph{network connection} to refer to a session where entanglement is established between a source and destination much like how network connections in traditional networks refer to a session where data is streamed between a source and destination. We say that a \emph{connection} corresponding to a network demand has been opened/closed when the quantum repeater protocols corresponding to that demand have been added/removed from the schedule.

\subsection{Schedule Overview}
To construct a schedule meeting QoS requirements, several actions are needed: First, the controller determines one (or several) options on how end-to-end entanglement generation can at all be realized such that the network demand submitted by two end nodes is satisfied.  This includes a path selection (see Routing) and a selection of repeater protocols for each path (see Protocol Selection). The choice of protocol determines which operations (see Fig.~\ref{fig:basics}) need to be executed at which nodes along the path, as well as the dependencies of said operations and timing constraints.

Second, given the selection of path and protocols, the central controller maps the protocol operations (and associated information) of all network demands into a joint network-wide schedule composed of fixed-duration time slots. Operations can span a single, or multiple consecutive slots, allowing additional time to be allocated to operations that require more time than others. Since operations occupy an integer number of slots, the size of slots should be chosen so that the excess amount of time allocated to operations is limited so as to limit reduction of fidelity from storing entangled links between operations.
%The slot size should additionally be upper-bounded by fixed-duration operations such as moving, entanglement swapping, and entanglement distillation.

Our scheduling structure is beneficial for several reasons. Flexibility in time allocation to operations allows us to spend appropriate amounts of time creating each elementary entangled link, which depends on the capabilities of the individual connected nodes as well as their distance in fiber (or free-space link). It also allows us to account for the highly varying capabilities in the different network nodes, where different platforms have different timing requirements and quality for the operations (see e.g.~\cite{humphreys2018deterministic,moehring2007entanglement}. These may even differ between two different nodes with the same underlying physical system due to the nature of early quantum hardware. Scheduling operations also ensures that qubits are exclusively allocated to each operation, granting contention-free usage of network devices to support network operation.
To guarantee consistent network operation, all quantum network nodes should be time-synchronized to boundaries of slots in the schedule. Such schedule synchronization can be achieved by building upon the existing synchronization mechanisms used by network nodes for heralded entanglement generation~\cite{dahlberg2019link}. 

Finally, the controller periodically installs the new schedule at all network nodes. New schedules can be installed in a synchronized fashion using a periodic reconfiguration period and having the controller instruct network nodes to switch to new schedules at pre-specified times. The period at which new schedules are installed is a design choice that depends on the desired responsiveness of the network and the minimum amount of time required to communicate the schedules to the network nodes. As time progresses, the schedule then directs network node behavior: it dictates when network nodes execute the operations in a repeater protocol.

\subsection{Constructing Schedules}
\label{sec:sched_constr}
We now detail the demand processing pipeline of our central controller as shown in Fig. \ref{fig:network_stack}.
We remark that in order to produce a schedule quickly and permit a modular design, we here separate the pipeline into distinct steps. Of course, one could envision that a global optimization could yield slightly better overall network performance, at the expense of a significantly higher computational effort increasing the latency.

\paragraph{Routing} The demand processing pipeline begins by first collecting demands submitted by network nodes. Depending on the desired responsiveness of the network, the central controller may be designed to process new network demands periodically or once a sufficient number of them have been aggregated. Collected demands are fed into a routing stage that determines paths of quantum network nodes to use for satisfying each network demand. At this stage, the central controller can assign paths to each network demand based on several different strategies. For example, paths may be assigned based on estimates of the achievable fidelity and throughput or in order to balance load across network resources. We remark that the problem of routing has been previously studied in several works \cite{chakraborty2019distributed,chakraborty2020entanglement,shi2020concurrent,pant2019routing} and that the choice of algorithm lies out of the scope of this paper.

\paragraph{Protocol Selection} After routing, quantum repeater protocols consisting of a combination of operations (see Fig.~\ref{fig:basics}) are chosen for each demand and associated path. This must take into account the capabilities of all the nodes along the paths (available qubits, quality of their memories and operations, time needed to execute operations), as well as the capabilities (fidelity, throughput, latency) for producing elementary links between connected nodes along the paths. The selection includes which type of qubits (communication or storage qubit) to use at which node. 
To provide an accurate estimate of the end-to-end fidelity, operations must be mapped onto qubits to determine 1) whether a sufficient number of qubits exist to execute the protocol, 2) determine any reduction in fidelity due to storage of entangled links between operations, and 3) the quality of the operations themselves. We define the latency of a quantum repeater protocol to be the amount of time that elapses between the start of the first operation and the completion of the final operation in the protocol, and we require that the latency of repeater protocols is small enough to meet throughput requirements. Mapping operations to qubits allows the controller to determine these quantities and find appropriate repeater protocols.  An example of such a mapping for the protocol in Fig. \ref{fig:example_rep_protocol} can be seen in Fig. \ref{fig:concrete_repeater_protocol_example}. Similarly to routing, protocol selection has been studied previously (\cite{KennethPaper,LukinPNAS,bacinoglu2010constant}, Appendix \ref{sec:app_rep_protocols}).

The specification of an operation in the schedule includes resource requirements (communication and storage qubits to be used), as well as QoS requirements (fidelity, throughput) of generating elementary links between connected nodes using a link layer protocol~\cite{dahlberg2019link}, allowing the nodes to correctly execute the operations.
Timing constraints take into account the memory quality of the network nodes involved, to ensure that entangled links are produced sufficiently close in time to allow for end-to-end entanglement generation fulfilling network demands within the limited memory lifetimes. 

Whenever an operation is probabilistic, sufficient time slots can in principle be allocated such that the operation(s) succeeds with high probability.
Entanglement generation of an elementary link can be achieved robustly using the link layer protocol in~\cite{dahlberg2019link}, where the distributed queue of~\cite{dahlberg2019link} is replaced by the schedule set by the controller. To abstract away from the details of a specific link layer protocol, we will below assume that a sufficiently large number of time is allocated in order to succeed almost surely in producing an elementary link within that time frame. In general, we remark our TDMA architecture can work hand in hand with the link layer protocol in~\cite{dahlberg2019link} where the granularity of time slots is chosen in a way that allow the link layer to work well with the physical layer. That is, the time slots are sufficiently large to allow for a practical batching of operations at the physical layer, but also not too large to lead to wasteful overheads which would be the case if indeed always slots were chosen to succeed with a heralding protocol for one pair. The latter is in general wasteful, since many more pairs are likely to be generated in consecutive slots that large, taking away a great amount of flexibility and efficiency at the link layer. We emphasize though that properly evaluating a more fine grained choice of slot size does require an evaluation of the total system (including the link layer and physical layer protocols), which is outside the scope of this paper where our focus is to take such protocols as abstract givens, and consider the general setup of a TDMA architeture and high level scheduling.

For larger end-to-end quantum repeater protocols, the probability that entanglement is successfully delivered depends on the creation of all elementary entangled links close in time. At a high level, one could use standard concentration bounds \cite{hoeffding1994probability} to determine the amount of time allocated to each elementary entangled link such that the overall protocol succeeds with high probability, although we remark that a choice of more fine grained slot sizes can be beneficial in practice, and calls for further research to be determined more generally.  Meeting throughput and jitter requirements becomes increasingly difficult if repeater protocols fail with non-negligible probability.

Despite scheduling probabilistic operations for sufficient time to succeed with high probability, entanglement generation may still fail. Such failures are communicated by the link and network layer to the end nodes, who then act in accordance with the application protocol to wait for the next link to succeed ($MD$ use cases, some $CK$ use cases, see Section \ref{sec:app_requirements}), or restart (most $CK$ use cases). No other action is taken and the nodes continue to follow the schedule.

\paragraph{Schedule Construction} The final stage of the central controller is to combine protocols for each pair of end nodes in order to construct the joint network schedule. Here, the chosen repeater protocols are scheduled so that the delivery of entanglement respects throughput and jitter requirements, where the controller may make a choice of different protocols identified to meet end node demands. This stage no longer considers the end-to-end fidelity requirements as the protocols have been chosen to satisfy minimum end-to-end fidelity. The schedule is constructed to be of a finite-length and is executed cyclically, meaning that the schedule repeats from the beginning once the end has been reached. By using a cyclic schedule, the central controller need only broadcast a new schedule to the network when demands are changed, thereby reducing the amount of communication to keep the network running. The length of the schedule is chosen by the controller based on the network demands. Once the schedule has been produced, the central controller distributes it to the network nodes and each node's reservation manager installs the schedule for local use. In Section \ref{sec:sched}, we will detail our approach to constructing these schedules.

When end nodes no longer require entanglement they contact the central controller to remove their network demands. If applications fail to remove their demands, the schedule will still retain the corresponding repeater protocol, potentially starving new applications. To avoid this, the central controller may employ a heartbeat mechanism where end nodes must regularly inform the central controller to keep their demand, allowing the controller to remove demands that it deems as inactive.

\section{Scheduling Repeater Protocols}
\label{sec:sched}
% The design considerations from section \ref{sec:des_con} pose several challenges in constructing TDMA schedules of repeater protocols.

As we have mentioned, the routing and protocol selection steps of the central controller's operation have previously been studied.  However, in order to realize our architecture, we require 1) a repeater protocol model that expresses resource and timing requirements, as well as 2) a method for producing a network-wide schedule from these representations and the network demands of end-nodes. In this section, we present our contribution for these requirements.  We will first show how we encode repeater protocols as directed,  acyclic graphs (DAGs) and then present two methods for constructing the network-wide schedule given a set of such DAGs along with their associated network demands.  While our network architecture supports both throughput and jitter requirements, in this section we focus on scheduling methods that fulfill throughput requirements while satisfying jitter requirements on a best-effort basis.

%We describe our model for the repeater protocols selected in the protocol selection stage and detail two methods for constructing the network schedule. A more in-depth overview of our methodology can be found in Appendix section \ref{sec:methdology}. 

%By encoding quantum repeater protocol operations into cyclic TDMA schedules, the network provides contention-free usage of network resources to deliver entanglement of desired fidelity at a frequency that respects end node's timing requirements on delivery. 

% GENERAL TEXT
% We first present our system models and the new problem of constructing TDMA schedules of quantum repeater protocols. We briefly describe two novel formulations of this scheduling problem under periodic task scheduling and resource-constrained project scheduling. 

% MENTIONED EARLIER
% Scheduling link establishment for entanglement swapping must be done such that both entangled links used in an entanglement swap are simultaneously available at repeater nodes. Furthermore, schedules should never attempt to produce additional links if there are no vacant qubits for holding the new entanglement. Respecting these constraints while also meeting varying levels of throughput, jitter, and fidelity requirements across network applications results in a complicated scheduling problem.

\subsection{Repeater Protocol Model}
\label{sec:rep_protocol_model}
The end-to-end fidelity of entanglement delivered by a repeater protocol depends on several factors such as the initial fidelity of entanglement produced between pairs of nodes as well as the amount of time entanglement is stored between entanglement swapping and distillation operations. It is in general a challenging question to select good repeater protocols (see e.g. \cite{azuma2021tools}) which is outside the scope of our work. Here, we take an existing repeater protocol as it may be found by algorithms such as \cite{azuma2021tools} as a given, and describe in detail how information made available by such repeater protocols can be represented in a form suitable for our scheduling methods.

The information made available by a repeater protocol should include a minimal fidelity for each elementary entangled link, the network nodes that perform each repeater protocol operation, the sequence of operations and their relative timing, as well as which qubits are used for each operation. Using these details, we develop a representation that allows an easy method for computing the worst-case end-to-end fidelity such that minimum fidelity requirements are expected to be met by the repeater protocol. To achieve this, we represent each repeater protocol as a directed acyclic graph (DAG) $P(A, I, M, Q)$.

In such a DAG, each vertex $a \in A$ represents an operation to perform.  Such an operation may be establishing entanglement between a pair of connected nodes with some initial fidelity, performing entanglement swapping of two entangled links, or executing entanglement distillation using two or more entangled links. Each operation $a \in A$ carries a tuple $(a_{ID}, a_V, a_F)$ where:
\begin{itemize}
	\item $a_{ID} \in \{link, swap, distill\}$ is an identifier of the type of operation to perform,
	\item $a_V$ is a set of network nodes that perform the operation, and
	\item for $ID=link$, $a_F$ specifies the initial fidelity of the elementary entangled link to produce.
\end{itemize}

Each edge $i \in I$ represents the dependency between two operations, where the dependency indicates that the entangled link produced from one operation is used as input to another operation.  We use $a \rightarrow b$ for $a,b \in A$ when operation $a$ has an outward edge to operation $b$, denoting that the link produced by $a$ is used as input to operation $b$. An operation $v \in A$ uses all entangled links produced by operations in the set $\{u \in A | u \rightarrow v\}$.  For example, an entanglement swapping operation will have two inward edges from operations that produced the entangled links to swap.  An entanglement distillation operation may have several inward edges, one for each entangled link to use for the operation, and several outward edges, one for each entangled link produced.  Elementary entanglement generation between pairs of nodes have no inward edges, and thus compose the set of sources in the DAG.

The relative offset map $M: A \rightarrow \mathbb{R} \times  \mathbb{R}$ maps each operation $a \in A$ to a start time $s_a$ and end time $e_a$ relative to the start of the protocol. This is needed as we wish to be able to express the latency of each operation in the repeater protocol when scheduling.  For example, entanglement generation between connected nodes that are farther apart will have higher latency than that between connected nodes that are closer together.  We may thus wish to allocate more time to the former in order to successfully create the entangled link.  A second instance of variable operation latency comes from entanglement distillation, where the required time to execute the operation will depend on the number of entangled links that are used and the number that need to be produced.  The relative offset map also provides us the latency of the overall repeater protocol and may be computed as $L_i=\max_{a \in A}e_a$.  For simplicity, we will assume that $s_a$ and $e_a$ for each operation are integer multiples of the slot size in the schedule.

Finally, the map $Q: A \rightarrow \mathbb{P}(K) \times \mathbb{P}(K)$ maps an operation $a$ to two sets of qubits belonging to the nodes in $a_V$.  The first set of qubits are those needed to perform the operation while the second set of qubits are those that hold an entangled link produced by the operation.  Here, the notation $\mathbb{P}(K)$ denotes the power set of all qubits in the network $K$.  We will frequently refer to the set $K$ as the set of network resources.   For an operation $a$ with $a_{ID}=link$, $Q(a)$ specifies which communication qubits should be used for establishing an entangled link with an adjacent network node and which qubits to store the created link in. $Q(a)$ for $a_{ID}=swap$ specifies the qubits holding entangled links to perform entanglement swapping with while $Q(a)$ for $a_{ID}=distill$ specifies the qubits holding links for entanglement distillation and the qubits to store the resulting links in.

To illustrate this representation, we present an example using the repeater protocol depicted in Fig. \ref{fig:example_rep_protocol}. This repeater protocol has three operations: two elementary link generation operations, $L_1$ and $L_2$, and one entanglement swapping operation, $S_1$.

The first elementary link generation, $L_1$, begins at time slot 0 and ends at time slot 2.  It uses three qubits: $A$-$comm.$ and $B$-$comm.$, which are communication qubits used to produce the entangled link between nodes $A$ and $B$, as well as $B$-$storage$, which is used by $B$ to store the created link and free $B$-$comm.$ to generate subsequent links. The second elementary link generation, $L_2$, begins after $L_1$ at time slot 2 and executes until the start of time slot 4.  This operation uses $B$-$comm.$ and $C$-$comm.$ to establish the next entangled link.  Since $B$-$storage$ is already holding the first link, we keep the second link in $B$-$comm.$. Finally, the entanglement swapping operation executes during time slot 4 and uses the entangled links stored in $B$-$comm.$ and $B$-$storage$ to produce an end-to-end link between $A$ and $C$.

\begin{figure}[h!]
	\centering
	\includegraphics[width=\linewidth]{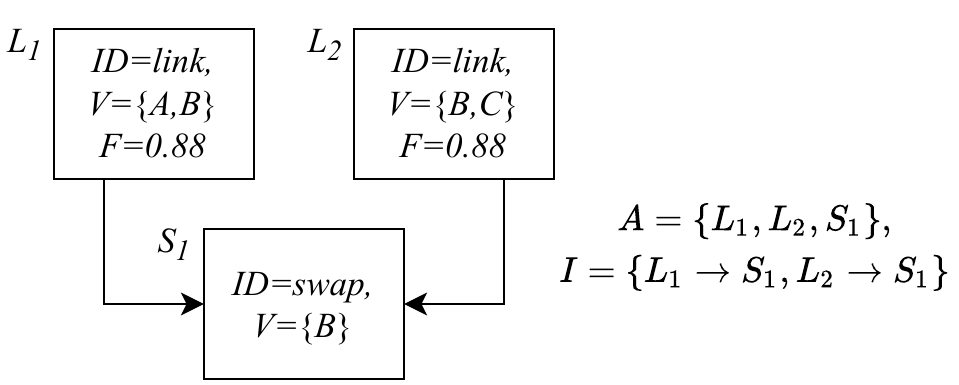}
	\caption{Example of a DAG representing the quantum repeater protocol used in Fig. \ref{fig:example_rep_protocol}.}
	\label{fig:repeater_protocol_dag}
\end{figure}

Fig. \ref{fig:repeater_protocol_dag} shows a DAG representing the set of operations, $A$, and operation dependencies, $I$, for this repeater protocol.  Here, we have specified that operations $L_1$ and $L_2$ should generate their elementary entangled links with fidelity $0.88$ as an example. From our description of the repeater protocol, $M$ and $Q$ are defined as:  

\[ M(a)=\begin{cases} 
(0 \textrm{ ms},20\textrm{ ms}) & a = L_1 \\
(20\textrm{ ms},40\textrm{ ms}) & a = L_2 \\
(40\textrm{ ms},50\textrm{ ms}) & a= S_1
\end{cases}
\]

\[ Q(a)=\begin{cases} 
(\{A\textrm{-}comm., B\textrm{-}comm., B\textrm{-}storage\},  & a = L_1 \\
\{A\textrm{-}comm., B\textrm{-}storage\}) \\
(\{B\textrm{-}comm., C\textrm{-}comm.\},  & a = L_2 \\
\{B\textrm{-}comm., C\textrm{-}comm.\}) \\
(\{B\textrm{-}comm., B\textrm{-}storage.\}, \{\}) & a= S_1
\end{cases}
\]

When using this representation, the worst-case end-to-end fidelity of a repeater protocol occurs when all entangled links are created at the start of their operation and their fidelity degrades while being stored between operations.  Since operations have pre-specified start and end times, the maximum storage time of each entangled link is known in advance.  An estimate of the worst-case end-to-end fidelity may then be computed directly using knowledge of operation and storage quality as is the case in our architecture's central controller. Alternatively, analytic formulae which depend on the storage times of entangled links may be used to calculate the fidelity\cite{vardoyan2021quantum}. We take this information here as a given, for example supplied by an existing algorithm as part of the selection of a quantum repeater protocol \cite{azuma2021tools}.

By modeling repeater protocols in this fashion, we may guarantee that end-to-end entanglement satisfies a minimum fidelity requirement by choosing repeater protocols that have a worst-case end-to-end fidelity satisfying the requirement.  Our approach to ensuring a worst-case end-to-end fidelity further demonstrates the complexity of scheduling repeater protocol operations.  Not only do we need to ensure that adjacent nodes in the network have available resources for generating elementary entangled links with one another, we additionally need to ensure that all nodes along the end-to-end path perform their operations in a timely manner such that any loss of fidelity from storing entanglement between operations still allows the repeater protocol to satisfy QoS requirements.  

This repeater protocol representation additionally lends itself to calculation of the worst-case probability of success.  Recall that elementary entangled link generation is probabilistic.  Given the duration of each elementary link generation operation, we may calculate the probability all links are successfully generated for the repeater protocol.  We may also calculate the success probability of entanglement distillation operations given knowledge of the worst-case fidelity of each link used for the operation. With this and knowledge of probabilistic entanglement swapping performed at network nodes, we may determine the overall worst-case success probability as the product of the success probabilities of all repeater protocol operations.

\begin{figure}[h!]
	\centering
	\includegraphics[width=0.75\linewidth]{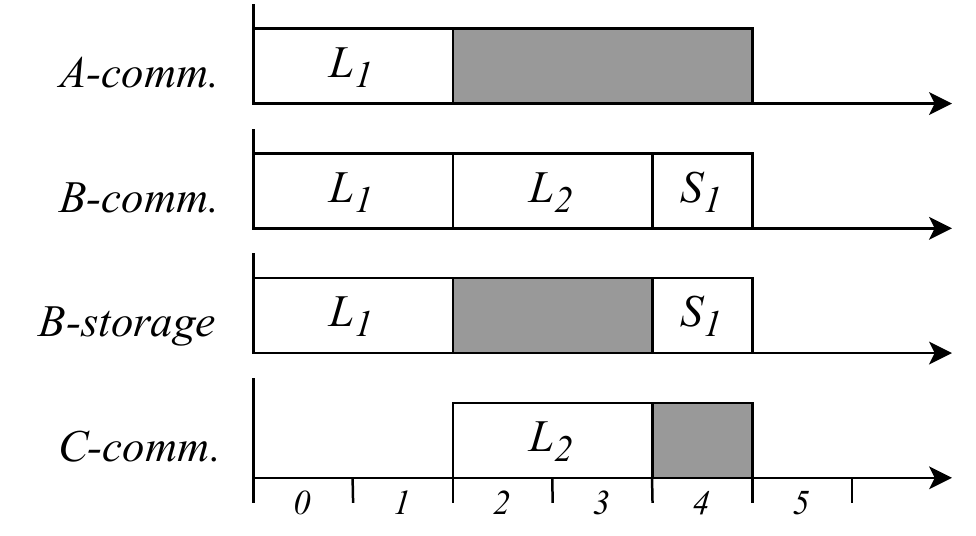}
	\caption{Schedule for the quantum repeater protocol in Fig. \ref{fig:example_rep_protocol} with operations $L_1,L_2$ (link generation) and $S_1$ (entanglement swapping). Communication and storage qubits for each node are labeled on the left while time proceeds to the right. Shaded regions indicate where qubits are holding entangled links.}
	\label{fig:concrete_repeater_protocol_example}
\end{figure}

Fig. \ref{fig:concrete_repeater_protocol_example} shows how our repeater protocol representation may be visualized in the form of a schedule timeline.  Here, we see how qubits across the network nodes are used for each repeater protocol operation.  In this figure, we additionally indicate regions of time where qubits are holding an entangled link and cannot be used in another operation.  We remark that alternatively one may specify the start and end times of only elementary link generation operations in such a schedule while move, entanglement swap, and entanglement distillation are assumed to be performed as soon as the required resources are available. In this case, the network layer would need to be informed of the protocol DAG to know how each elementary entangled link is used to deliver end-to-end entanglement.

Throughout the remainder of this paper, we assume that the protocol selection stage of our central controller provides the chosen repeater protocols for each demand in this representation to the TDMA schedule construction stage.  Furthermore, we additionally assume that these repeater protocols each have a worst-case end-to-end fidelity meeting the QoS requirements of network demands and succeed with high probability.

\subsection{Scheduling Problem}
\label{sec:sched_prob}
In our architecture, the goal of the TDMA schedule construction step is to produce a network-wide schedule of repeater protocol operations such that entanglement is delivered to pairs of users in the network according to their minimum fidelity and rate requirements as well as their maximum jitter requirements.

The problem of constructing TDMA schedules of quantum repeater protocols can be stated as follows. Given a set of network demands, $\mathcal{D}$, containing $(src,dst,F_{min},R_{min},J_{max})$ tuples detailing the source, destination, minimum fidelity, minimum throughput, and maximum jitter requirements of each demand along with a corresponding set of repeater protocols $\mathcal{P}$ containing $P(A, I, M, Q)$ DAGs for each demand, produce a schedule $\mathcal{S}$ that maps each (demand, protocol) pair $(D_i, P_i)$ to a set of start times $\mathcal{S}(D_i, P_i)=\{s_1,...,s_k\}$ such that

%that maps operations $a \in A_i$ of each concrete repeater protocol $P_i(A_i,I_i,M_i,Q_i)$ to a set of start times \textbf{$s_{a}$}$=\{s_{a,1},...,s_{a,k}\}$ such that:

\begin{enumerate}
	\item the end-to-end entanglement delivered by each protocol $P_i$ is at least $F_{min}^i$,
	\item the average rate of entanglement delivery for demand $D_i$ satisfies minimum rate requirements,
	$$\frac{|\mathcal{S}(D_i, P_i)|}{L} \ge R_{min,i},$$and
	\item the variance in inter-delivery times for a demand $D_i$ is below $J_{max}^i$, 
	$$\frac{\sum_{j=1}^{|\mathcal{S}(D_i, P_i)|} (s_{j+1}-s_j - \hat{s}_i)^2}{|\mathcal{S}(D_i, P_i)|} \le J_{max,i},$$
\end{enumerate}

where $L$ is the duration of the schedule, the quantity $s_{j+1} - s_j$ is the inter-delivery time between the $j$th and $(j+1)$th delivery for demand $D_i$, and $\hat{s}_i$ is the average inter-delivery time of $D_i$.

From our repeater protocol representation, we know that if each repeater protocol $P_i$ chosen by the protocol selection stage has a worst-case end-to-end fidelity larger than $F_{min}^i$, then we will satisfy the first requirement as long as we schedule its operations according to the maps $M_i$ and $Q_i$ describing the timing and resource requirements of the protocol.  To ensure this, the following constraints are placed on the scheduling problem:

\begin{enumerate}
	\item operations in $A_i$ are scheduled to respect the relative offset mapping $M_i$ of repeater protocol $P_i(A_i,I_i,M_i,Q_i)$. That is, for any operations $a,b \in A_i$, the $j$th start times $s_{a,j},s_{b,j}$ satisfy $s_{a,j}-s_{b,j} = M_i(a) - M_i(b)$ for all $i$ and $j$,
	\item no two operations $a\in A_i$, $b \in A_j$ of two repeater protocols $P_i$ and $P_j$ that operate on some mutual set of qubits $Q_i(a) \cap Q_j(b) \neq \emptyset$ are assigned start times $s_{a,j}, s_{b,k}$ and end times $e_{a,j}, e_{b,k}$ such that $s_{a,j} \leq s_{b,k} < e_{a,j}$ or $s_{b,k} \leq s_{a,j} < e_{b,k}$,
	\item no distinct operations $a,b,c \in \cup_{P_i \in \mathcal{P}} A_i$ with $b \rightarrow c$ are assigned start times $s_{a,j},s_{b,k},s_{c,k}$ such that $s_{b,k} \leq s_{a,j} \leq s_{c,k}$ and $Q(a) \cap Q(b) \cap Q(c) \neq \emptyset$ for any $j,k$,
\end{enumerate}

\noindent where the first condition ensures we respect the timing requirements of the repeater protocol, the second condition ensures that operations are granted exclusive access to their required qubits, and the third condition ensures no operation attempts to use a qubit that holds an entangled link intended for a different operation. Note that our definition of QoS will be satisfied with high probability provided the protocol selection stage chooses repeater protocols with high success probability. The case where entanglement delivery always succeeds may be recovered by delivering a classically-correlated quantum state to applications such that the average entanglement fidelity satisfies fidelity requirements \cite{humphreys2018deterministic}.

Consider the schedule shown in Fig. \ref{fig:concrete_repeater_protocol_example} and suppose it has a duration $L$ of $50$ ms (5 time slots), each of fixed-duration $10$ ms, after which the schedule executes cyclically from slot 0.  In this instance, we deliver an end-to-end entangled link every $50$ ms, giving us a rate of $20 \frac{ebit}{s}$ and a jitter of $0$ $s^2$.  Assuming the encoded repeater protocol has a worst-case end-to-end fidelity that satisfies the minimum fidelity requirement, this schedule may be used for any rate requirement $R_{min} \leq 20\frac{ebit}{s}$ and any jitter requirement $J_{max}$.

Now suppose node $C$ is connected to another node $D$ and they also wish for the network to deliver an entangled link between them at a rate of $20\frac{ebit}{s}$.  Suppose further that their repeater protocol consists of a single elementary link generation operation, $L'_1$, which requires one time slot ($10$ ms) to succeed with high probability.  Fig. \ref{fig:two_protocol_mapping} depicts a valid and invalid scheduling of $L'_1$ into the existing schedule that meets $A$ and $C$'s requirement.  Here, making the schedule 6 slots long by scheduling $L'_1$ at slot 5 results in a rate of $16.67 \frac{ebit}{s}$ for both pairs $(A,C)$ and $(C,D)$, which fails to meet QoS requirements while scheduling $L'_1$ at time slot 0 permits both pairs of nodes to receive a throughput of $20\frac{ebit}{s}$.

\begin{figure}[h!]
	\centering
	\includegraphics[width=0.75\linewidth]{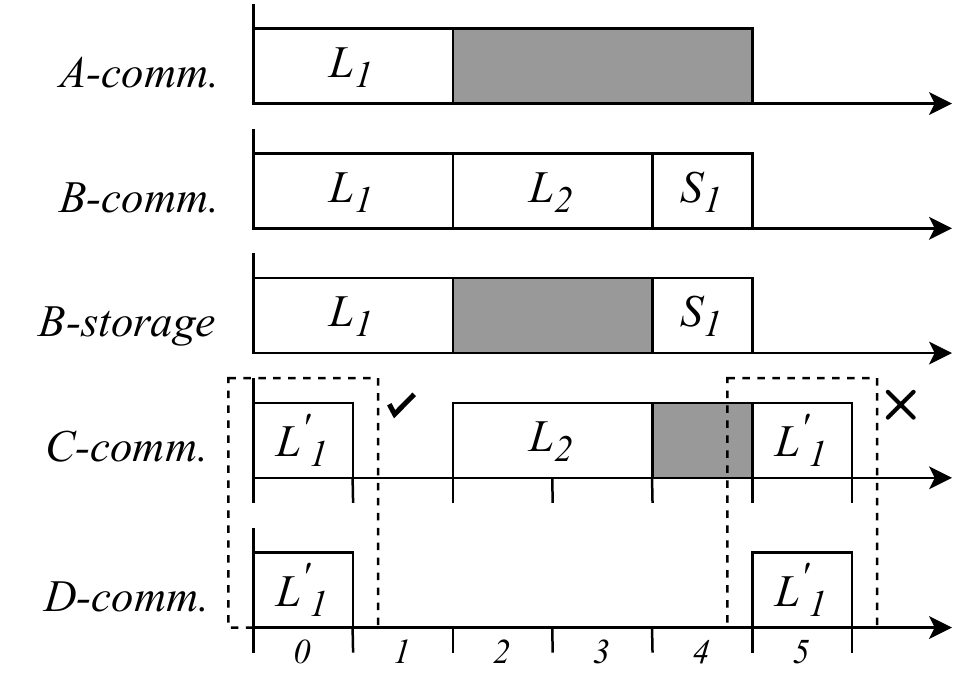}
	\caption{Example of a schedule with a valid and invalid placement of repeater protocol operation $L'_1$.}
	\label{fig:two_protocol_mapping}
\end{figure}

\subsection{Scheduling Methods}
\label{sec:sched_med}
In this section, we will detail our contribution on how we approach the schedule construction step using two different methods based on periodic task scheduling and resource-constrained project scheduling. While our architecture is suitable for handling both throughput and jitter QoS requirements, we limit the scope of our scheduling methods to handle throughput requirements while fulfilling jitter requirements on a best-effort basis. We remark however that the methods we present here provide an upper-bound on the observed jitter $J_i$ for a given demand $D_i$ of $\frac{1}{R_{min,i}}^2$.  We also present extensions to these scheduling methods in order to handle general jitter constraints in Appendix \ref{sec:app_sched_ext}.

Our scheduling methods are motivated by the real-time constraints imposed by near-term quantum hardware and we propose the use of a novel heuristic that combines the two scheduling methods presented below. We benchmark our heuristic against other known heuristics from these scheduling methods in section \ref{sec:sched_eval}.

\subsubsection{Periodic Task Scheduling}
\label{sec:pts}
Given our problem definition, the first method we consider for constructing the schedule is through non-preemptive periodic task scheduling techniques \cite{codd1960multiprogram,liu1973scheduling}.  Originally studied in the context of real-time systems, periodic task scheduling is the problem of scheduling a set of tasks for execution on a processor.  Each task is released into the system according to a fixed period and must execute for some worst-case execution time exclusively on the processor in order to complete. The goal is to schedule each task instance released into the system for execution such that it completes before the end of the period it was released.  In the non-preemptive case, tasks that start execution on the processor must run to completion.

To use this method for our scheduling problem, each repeater protocol DAG is transformed into a periodic task with execution requirements reflecting the protocol's latency and the QoS requirements of the corresponding demand. Simulating the scheduling of the periodic task set produces a schedule indicating when each repeater protocol starts in the network-wide schedule.

We first describe abstractly the elements to be defined in the problem of periodic task scheduling before explaining the mapping of our problem to it.

\begin{itemize}
	\item $\tau_i=(\Phi_i, C_i, T_i)$: The $i$th periodic task, defined by three parameters: The release offset $\Phi_i$ (set to 0), the worst-case execution time $C_i$, and the period $T_i$.
	\item $\tau_{i,j}$: The $j$th instance of task $\tau_i$.
	\item $C_i$: The worst-case execution time of $i$th task. Each instance $\tau_{i,j}$ of $\tau_i$ has this worst-case execution time.
	\item $T_i$: The period of task $\tau_i$, specifies the amount of time between subsequent releases of instances of $\tau_i$.
	\item $\Gamma$: A set of tasks.
	\item $H$: The hyperperiod of the schedule, computed as the least common multiple (LCM) of the task periods \\ $LCM(T_1, ... , T_n)$.
\end{itemize}

Determining a scheduling of the periodic task set within the time interval $[0,H]$ provides a scheduling for each interval $[kH, (k+1)H]$ due to the fact that tasks are released into the system according to the same pattern as in $[0,H]$.  This property allows us to find a cyclic schedule by finding a scheduling up to the hyperperiod.

Given a set of network demands $\mathcal{D}$ and their corresponding repeater protocols $\mathcal{P}$, the periodic task set $\Gamma$ is constructed as follows. For each demand $(src_i,dst_i,F_{min,i},R_{min,i}) \in \mathcal{D}$ and associated repeater protocol $P_i(A_i, I_i, M_i, Q_i)$, a periodic task $\tau_i$ is constructed with phase $\Phi_i=0$, worst-case execution time $C_i=\frac{L_i}{t_{slot}}$ (the repeater protocol latency in slots), and period $T_i=\lfloor \frac{1}{t_{slot} R_{min,i}} \rfloor$. $C_i$ is chosen to reflect the duration of the repeater protocol while $T_i$ is chosen to respect the rate requirement of the demand.  $t_{slot}$ is the duration of a time slot in the schedule and $t_{slot}R_{min,i}$ must be less than 1, otherwise no periodic task schedule can satisfy the rate requirement $R_{min,i}$. Choosing $\Phi_i=0$ minimizes the length of the schedule and allows the hyperperiod $H$ to be pre-computed by the least common multiple (LCM) of the task periods $LCM(T_1, ... , T_n)$.

Solving the resulting scheduling problem satisfies rate requirements for the following reasons.  Each periodic task $\tau_i$ corresponding to repeater protocol $P_i$ executes a total of $X_i=\frac{H}{T_i}$ times over the course of a schedule of length $H$ slots.  Choosing $T_i=\lfloor \frac{1}{ t_{slot}R_{min,i}}\rfloor$ guarantees the observed rate $R_i$ of entanglement delivery for demand $D_i$ satisfies a throughput requirement of $R_{min,i}$ due to the fact that

$$R_i=\frac{X_i}{Ht_{slot}}=\frac{1}{T_i t_{slot}}=\frac{1}{\lfloor \frac{1}{t_{slot} R_{min,i}} \rfloor t_{slot}} \ge R_{min,i}.$$

For jitter requirements, the periodic task scheduling method provides an upper-bound on the observed jitter $J_i$ for a demand $D_i$ of $\frac{1}{R_{min,i}}^2$ as

\begin{align*}
J_i &= \frac{\sum_{j=1}^{|\mathcal{S}(D_i, P_i)|} (s_{j+1} - s_j - \bar{s}_i)^2}{|\mathcal{S}(D_i, P_i)|}\\
&\le \frac{|\mathcal{S}(D_i, P_i)|(T_it_{slot})^2}{|\mathcal{S}(D_i, P_i)|} = (\frac{1}{R_i})^2 \\
&\le \frac{1}{R_{min,i}}^2.
\end{align*}

\noindent Here, we have made use of the fact that  $L_i \leq s_{j+1}-s_j \leq 2T_it_{slot}-L_i$ and that

$$\bar{s}_i=\frac{\sum_{j=1}^{|\mathcal{S}(D_i, P_i)|}(s_{j+1}-s_j)}{|\mathcal{S}(D_i, P_i)|}=\frac{Ht_{slot}}{|\mathcal{S}(D_i, P_i)|}=T_it_{slot}$$

\noindent which allows us to bound the quantity $(s_{j+1} - s_j - \bar{s}_{i,j})$ by

\begin{align*}
L_i - T_it_{slot} \le (s_{j+1} - s_j - \bar{s}_{i,j}) \le T_it_{slot} - L_i.
\end{align*}

\noindent Using the fact that $T_i$ must be larger than $C_i=\frac{L_i}{t_{slot}}$ in order for a task instance $\tau_{i,j}$ to have enough time to complete within the period it was released, we obtain the bound $(s_{j+1} - s_j - \bar{s}_{i,j})^2 \le (T_it_{slot} - L_i) < (T_it_{slot})^2$.

Consider our previous example from Fig. \ref{fig:two_protocol_mapping} using a time slot size of $10$ ms and suppose a rate of $16\frac{ebit}{s}$ is acceptable between $(A,C)$ and $(C,D)$.  Repeater protocol $P_1$ between $(A,C)$ is transformed into a periodic task $\tau_1$ with $\Phi_1=0$, $C_1=5$, and $T_1=6$ while repeater protocol $P_2$ between $(C,D)$ is transformed into a periodic task $\tau_2$ with $\Phi_2=0$, $C_2=1$, $T_2=6$.  The hyperperiod $H=LCM(T_1,T_2)=6$, so we only need a schedule of 6 slots which executes cyclically and we only need one instance $\tau_{1,1},\tau_{2,1}$ of each repeater protocol in the schedule.  Two valid schedules produced by this method are shown in Fig. \ref{fig:pts_example}.

\begin{figure}[h!]
	\centering
	\includegraphics[width=0.75\linewidth]{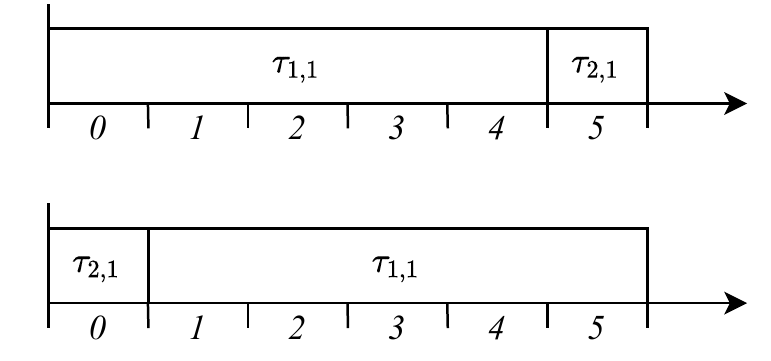}
	\caption{Two periodic task schedules that satisfy a rate requirement of $16\frac{ebit}{s}$ on two repeater protocols $P_1$ and $P_2$. Here, each time slot has a fixed-duration of $10$ ms.}
	\label{fig:pts_example}
\end{figure}

We choose to use non-preemptive scheduling due to two reasons: 1) permitting preemption of a protocol mid-execution introduces delays between operations that increase the latency of the quantum repeater protocol and also reduce the worst-case end-to-end fidelity, preventing it from meeting QoS requirements and 2) qubits may hold entangled links at the time of preemption, meaning they are either unavailable for use by the preempting protocol or the entangled links must be discarded, resulting in failure of the previously executing protocol.  By determining a schedule for the set of periodic tasks, a corresponding network schedule can be extracted that specifies when each repeater protocol $P_i(A_i,I_i,M_i,Q_i)$ starts. The start and end times for each repeater protocol's operations are then obtained by using the associated map $M_i$.

We remark that the use of preemptive strategies can offer higher scheduling flexibility and permit additional demands to be satisfied, though this comes at the cost of ensuring that the entanglement created before the period of preemption is still present in order to resume the protocol afterwards.  Furthermore, preemption of repeater protocols introduces delays between operations which may not be permitted due to memory lifetimes and can reduce the worst-case end-to-end fidelity below its acceptable requirement.  The worst-case end-to-end fidelity of a repeater protocol must then compensate for any introduced delays due to preemption.

In general, determining a valid schedule for a set of non-preemptable periodic tasks is NP-hard \cite{jeffay1988optimal}, thus the choice of algorithm impacts the overhead in producing a new schedule and consequently the network's responsiveness to changes in network demands. To achieve lower overhead, scheduling heuristics can be used to construct schedules more quickly than an exhaustive search at the cost of not scheduling some of the tasks. With respect to the class of work-conserving scheduling techniques, non-preemptive earliest-deadline first (NP-EDF) \cite{jeffay1991non} is known to be optimal, meaning that NP-EDF can schedule a set of tasks if there exists a non-preemptive work-conserving heuristic capable of scheduling the same set of tasks. Our implementation of NP-EDF has a runtime complexity of $O(N\log N)$ where $N$ is the number of scheduling decisions made and

$$N = \sum_1^n \frac{H}{T_i}$$

\noindent due to the fact we only construct the schedule up to the hyperperiod $H$.

While the periodic task scheduling method is simple, it comes at the cost of lower network throughput. In our previous example, we would not be able to use this technique to meet fidelity requirements of $20\frac{ebit}{s}$ between pairs $(A,C)$ and $(C,D)$. This is due to the fact that fine-grained scheduling decisions on individual qubits are hidden in order to reduce the complexity of creating the schedule.  Under high network load this results in treating the network as a single resource which prevents any concurrent execution of quantum repeater protocols that operate on disjoint paths in the network.

One may try to mitigate this effect by preprocessing the set of repeater protocols $\mathcal{P}$ into disjoint subsets $\mathcal{P}_1,...,\mathcal{P}_k$ operating on disjoint sets of network resources.  Since no two repeater protocols $P_i \in \mathcal{P}_i,P_j \in \mathcal{P}_j$ require the same qubits, they may be scheduled independently of one another.  A set of periodic tasks $\Gamma_i$ is created for each subset of repeater protocols $\mathcal{P}_i$ and a network-wide schedule may be found by merging the schedule computed for each subset.  Determining the set of disjoint subsets can be achieved using a path-vertex intersection graph where vertices represent repeater protocols and edges represent pairs of repeater protocols that share one or more qubits as required resources. The set of vertices in each disjoint component of the path-vertex intersection graph then corresponds to a subset of repeater protocols $\mathcal{P}_i$ that share network resources and must be scheduled as part of the same task set $\Gamma_i$. 

This preprocessing step only remedies the case where network load is low and independent sets of resources can be assigned to each repeater protocol. We next describe a second method that is able to address this and achieve higher network performance at the cost of additional complexity in schedule construction.

\subsubsection{Resource-Constrained Project Scheduling}
\label{sec:sched_rcpsp}
Creating the schedule can be achieved using a second method based on the non-preemptive resource-constrained project scheduling problem (RCPSP) \cite{brucker1999resource}. Originally studied in the context of operations research, the goal of RCPSP is to schedule activities of a project under scarce resource constraints and precedence relations. Frequently, a project is represented as an activity-on-node network where each activity specifies a required set of resources and a duration for which it must be executed uninterrupted.  The set of resources in RCPSP is renewable, meaning that a resource needed for an activity becomes free once an earlier activity has completed.  Precedence constraints in the form of minimal and maximal time lags indicate the earliest and latest point an activity must begin processing after an earlier activity has completed.    

To apply this method to our scheduling problem, quantum repeater protocol operations are encoded into the activities of an activity-on-node network representing a project for RCPSP. Resource and precedence constraints are derived from the the resource map $Q$ and relative offset mapping $M$ for each quantum repeater protocol.  Scheduling the activity-on-node network provides a scheduling of all quantum repeater protocol operations which may be used to determine the network-wide schedule. 

We first present a definition of the notation we use here for RCPSP.
\begin{itemize}
	%	\item $J$: The set of all activities in the project.
	\item $j_i$: An activity in the project.
	\item $j_s, j_e$: Dummy start and end activities in an activity-on-node network. Has zero processing time and no resource requirements.
	\item $p_i$: Processing time of activity $j_i$.
	\item $K$: The set of all resources.
	\item $h_{ik}$: Required number of resource $k$ by activity $j_i$ to execute. $j_i$ may not begin unless $h_{ik}$ units of resource $k$ are available.
	\item $d_{ij}$: Minimal time lag between activities $j_i$ and $j_j$. Constraints the starting time of $j_j$ to be at least $d_{ij}$ units after the completion time of $j_i$.
	\item $\bar{d}_{ij}$: Maximal time lag between activities $j_i$ and $j_j$. Constraints the starting time of $j_j$ to be at most $\bar{d}_{ij}$ units after the completion time of $j_i$.
\end{itemize}

The RCPSP method first constructs an instance of the activity-on-node network for each repeater protocol $P_i(A_i,I_i,M_i,Q_i)$ in $\mathcal{P}$. At this level, minimal/maximal time lags between activities are chosen to respect the relative time mapping $M_i$ the project activities themselves are non-preemptable to respect the estimated end-to-end fidelity of the repeater protocol. The set of resources $K$ is the set of communication and storage qubits present at the nodes in the network. First, we show how we construct the activity-on-node network for a single repeater protocol and then show how the full activity-on-node network is constructed for the scheduling problem.

Producing an instance of the activity-on-node network for a repeater protocol $P(A,I,M,Q)$ begins by constructing dummy start and end activities $j_s$ and $j_e$ with processing times $p_s=p_e=0$ and $h_{sk}=h_{ek}=0$ for all $k \in K$. A project activity $j_a$ is constructed for each repeater protocol activity $a \in A$ with $M(a)=(s_a,e_a)$ having processing time $p_a=\frac{e_a-s_a}{t_{slot}}$ and resource requirements $h_{ak}=1$ for all $k \in Q(a)$.  An edge is then created between $j_s$ and each such activity $j_a$ with $d_{sa}=\bar{d}_{sa}=\frac{s_a}{t_{slot}}$. Next, for each resource $k$ we create a list of activities $A_k$ that use $k$ sorted by their starting times.  For each pair of subsequent operations $a,b \in A_k$ with start and end times $s_a,e_a$ and $s_b,e_b$ respectively, if operation $a$ stores an entangled link in $k$ then we construct an activity $j_o$ with processing time $p_o=\frac{s_b-e_a}{t_{slot}}$, resource requirement $h_{ok}=1$, and minimal/maximal time lags $d_{ao}=\bar{d}_{ao}=d_{ob}=\bar{d}_{ob}=0$.  If the last operation $a$ in this list stores an entangled link in $k$, then we create an activity $j_o$ with processing time $p_o=\frac{L_i-e_a}{t_{slot}}$, resource requirement $h_{ok}=1$, and minimal/maximal time lags $d_{ao}=\bar{d}_{ao}=0$ between $j_a$ and $j_o$ as well as minimal/maximal time lags $d_{oe}=\bar{d}_{oe}=0$ between $j_o$ and the dummy end activity $j_e$.  Finally, for all $a \in A$ with end time $e_a=L_i$, we specify minimal/maximal time lags $d_{ae}=\bar{d}_{ae}=0$.

As an example, let us consider the repeater protocols from Fig. \ref{fig:two_protocol_mapping}.  The activity-on-node network for each repeater protocol may be visualized in Fig. \ref{fig:aon_example}. Here, "$O$" nodes are activities that representing occupation of resources between operations.

\begin{figure}[h!]
	\centering
	\begin{subfigure}{\linewidth}

		\caption{}
		\includegraphics[width=\linewidth]{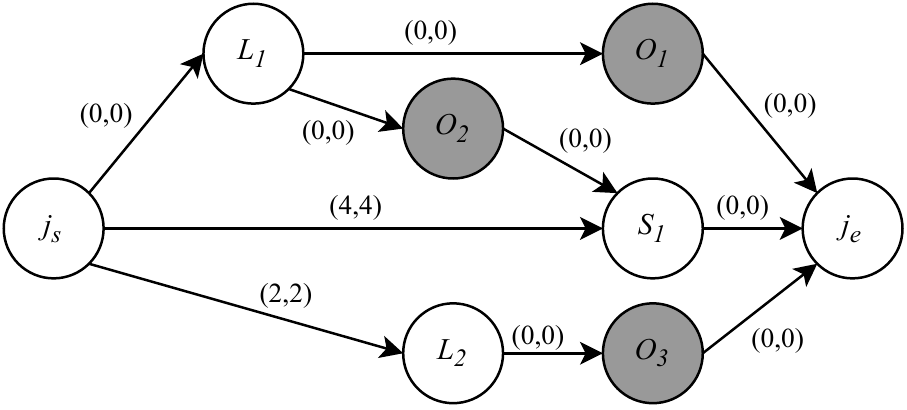}
	\end{subfigure}

	\begin{subfigure}{\linewidth}
		\centering
		\caption{}
		\includegraphics[width=0.55\linewidth]{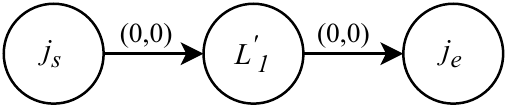}
	\end{subfigure}
	\caption{Activity-on-node networks for the repeater protocols used for demands between a) $(A,C)$ and b) $(C,D)$.}
	\label{fig:aon_example}
\end{figure}

Our approach for constructing the full activity-on-node network for the scheduling problem reuses the notion of a hyperperiod from periodic task scheduling in order to satisfy throughput requirements.  First, we create a new pair of dummy start and end activities $j_s$ and $j_e$. A period $T_i=\lfloor \frac{1}{t_{slot} R_{min,i}} \rfloor$ is computed for each demand $D_i$ and the hyperperiod is computed similarly as $H=LCM(T_1,...,T_n)$.  We then create and enumerate $X_i=\frac{H}{T_i}$ instances of the activity-on-node network for repeater protocol $P_i$ and specify minimal/maximal time lags between the dummy start activities $j_s$ and $j_{s_l}$ for the $l$th instance as $d_{ss_l}=lT_i$ and $\bar{d}_{ss_l}=lT_i-\frac{L_i}{t_{slot}}$ for $0 \leq l < X_i$.  Finally, we specify minimal/maximal time lags between the $l$th instance's dummy end $j_{e_l}$ and the dummy end $j_e$ as $d_{e_l e}=0$ and $\bar{d}_{e_le}=T_i-\frac{L_i}{t_{slot}}$.

The additional minimal and maximal time lags specified between the dummy start $j_s$ and each instance's dummy start $j_{s_l}$ as well as each instance's dummy end $j_{e_l}$ and the dummy end $j_e$ restrict the $l$th instance to execute within the period $[lT_i,(l+1)T_i]$ similarly to how tasks are scheduled in periodic task scheduling. Due to this, our RCPSP method satisfies minimum rate requirements and provides an upper-bound of $\frac{1}{R_{min,i}}^2$ as well.

\begin{figure}[h!]
	\centering
	\includegraphics[width=\linewidth]{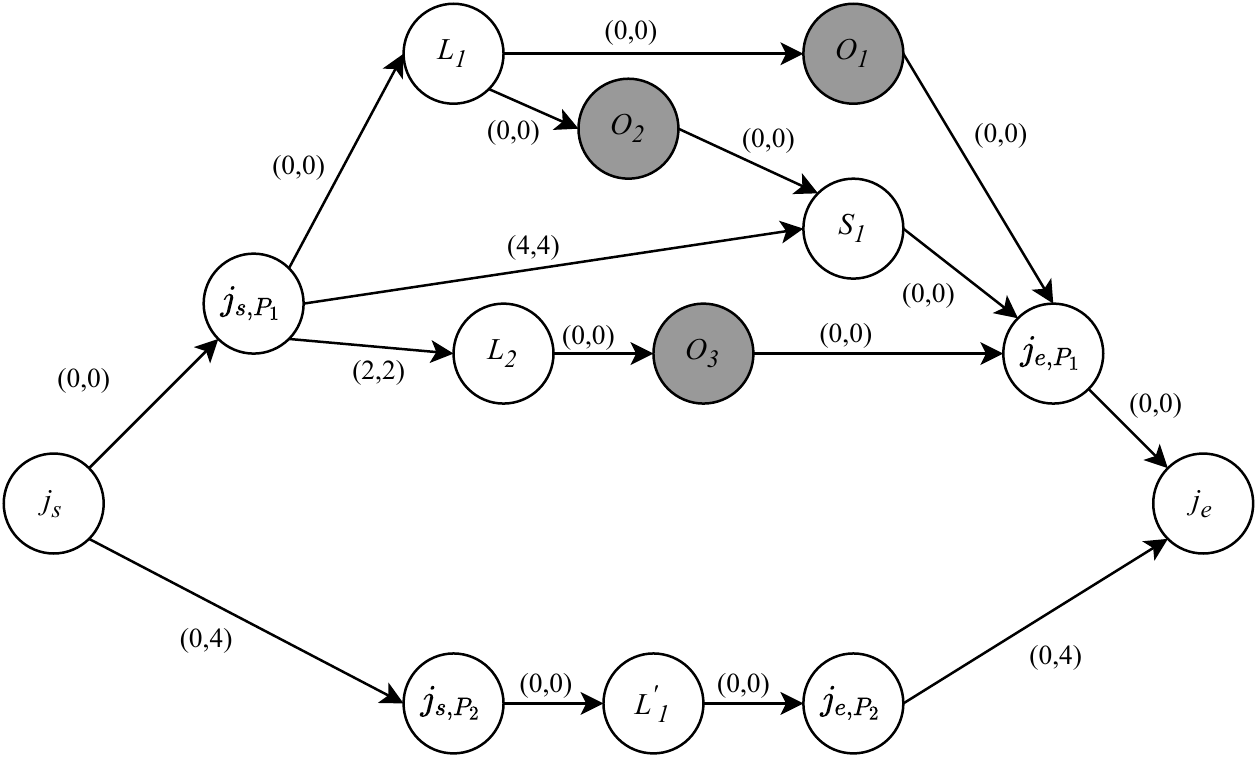}
	\caption{Full activity-on-node network for the example repeater protocols in Fig. \ref{fig:aon_example}. Here, the dummy start and end activities for repeater protocols $P_1$ and $P_2$ have been relabeled as $j_{s,P_1},j_{e,P_1}$ and $j_{s,P_2},j_{e,P_2}$ for clarity.}
	\label{fig:aon_example_full}
\end{figure}

A scheduling of the full activity-on-node network provides the start times of each repeater protocol operation and consequently the start time of each protocol.  Since the activity-on-node networks for each repeater protocol instance are constructed reflecting the resource and timing requirements of each protocol, we know that we satisfy the conditions set out previously. The activity-on-node network for our example can be visualized in Fig. \ref{fig:aon_example_full}.

The number of activities in the full activity-on-node network scales as $O(NS)$ where $N$ is the total number of repeater protocol instances encoded in the final network and $S$ is the maximum number of operations in an instance of a repeater protocol. This results from each activity in the activity-on-node network corresponding to an operation for an instance of a repeater protocol in the network. Since we construct multiple instances of each repeater protocol based on the rate it must deliver entanglement, a repeater protocol's operations may appear multiple times in the final network. 

Similarly to periodic task scheduling, constructing a schedule in RCPSP is NP-hard \cite{bartusch1988scheduling} and heuristic methods can be used to find schedules more quickly \cite{neumann1999methods}. Due to the added complexity from scheduling individual resources, RCPSP heuristic solvers observe a higher complexity than periodic task scheduling heuristic algorithms. The trade-off with this higher complexity is that using RCPSP to represent the scheduling problem allows finer-grained scheduling decisions at the resource level, permitting higher levels of parallelism between repeater protocol operations.

We evaluate two heuristics for the RCPSP scheduling method. The first heuristic, referred to as RCPSP-NP-EDF, is based on the NP-EDF heuristic from Section \ref{sec:pts} while the second heuristic combines the periodic task scheduling and RCPSP methods together into a heuristic we call full-protocol reservation (RCPSP-NP-FPR). This novel heuristic approximates the activity-on-node network to a fixed-size that is independent of the number of operations in the protocol. Doing so provides a reduction in runtime complexity compared to the RCPSP-NP-EDF heuristic while still achieving higher network performance than periodic task scheduling as we show in section \ref{sec:eval}.  

Our heuristic works as follows: we replace each instance of the activity-on-node network for a repeater protocol with one composed of three activities, $j_s,j_a,j_e$, where $j_s$ and $j_e$ are the dummy start and end activities as before, but $j_a$ is an activity with processing time $p_a=\frac{L_i}{t_{slot}}$, resource requirements $h_{ak}=1$ for all $k \in \cup_{a \in A}Q(a)$, and minimal/maximal time lags $d_{sa}=\bar{d}_{sa}=d_{ae}=\bar{d}_{ae}=0$.  In essence, the entire repeater protocol's latency and resource requirements are encoded into a single activity that requires all qubits for the full duration of the protocol.  The scheduled start time of each activity $j_a$ corresponds to a start time of the repeater protocol it represents.  As an example, we show the modified activity-on-node network for our previous example in Fig. \ref{fig:aon_example_full_fpr}.

\begin{figure}[h!]
	\centering
	\includegraphics[width=\linewidth]{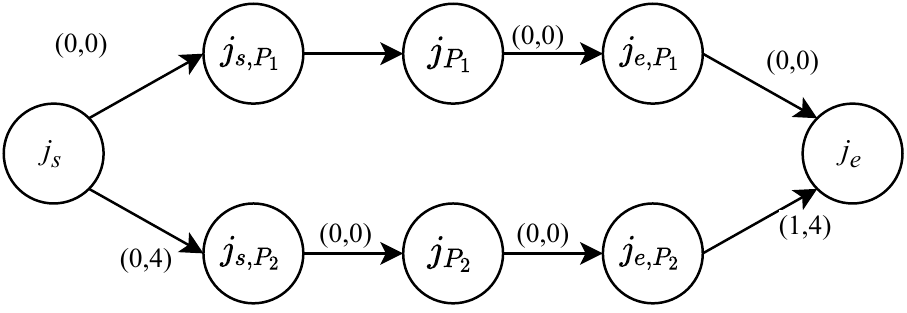}
	\caption{Modified activity-on-node network for the example in Fig. \ref{fig:aon_example_full}.  The activity-on-node network instance for repeater protcol $P_1$ has been condensed into a three-activity network.}
	\label{fig:aon_example_full_fpr}
\end{figure}

Our RCPSP-NP-EDF and RCPSP-NP-FPR heuristics have runtime complexities $O(N^2S^2|K|\log(NS))$ and $O(N^2|K|\log(N))$ respectively where $N$ is the same as in the periodic task scheduling method, $|K|$ is the total number of qubits, and $S$ is the maximum number of operations in any repeater protocol to schedule. $S$ depends on the quality of quantum operations and memory lifetimes as well as the number of hops between the pairs of nodes and the desired end-to-end fidelity, making it difficult to approximate.

% GENERAL INFO
% When there is high network congestion, the periodic task scheduling formulation causes scheduling decisions to treat large portions of the network as a single resource, meaning that protocols will be scheduled serially when they can be performed concurrently. The RCPSP formulation resolves this by performing scheduling decisions on a per-network resource basis, permitting concurrent execution of quantum repeater protocols.

%\begin{figure*}[htb!]
%	\centering
%	\includegraphics[width=\linewidth]{figures/latency_vs_slot_size.png}
%	\caption{Latency and the required number of consecutive slots (Num Slots) to encode quantum repeater protocols for two directly connected nodes and a two hop path for fidelity requirements F. Network devices have one communication qubit and three storage qubits. Larger values of F correspond to higher quality entanglement. }
%	\label{fig:latency_v_slot_size}
%\end{figure*}

\section{Evaluation}
\label{sec:eval}
At present, the largest network of processing nodes consists of three nodes~\cite{upcomingGHZ}, prohibiting an interesting hardware validation. We instead focus on an evaluation in simulation using some hardware validated parameters. The objective of our architecture is to enable such early stage networks to scale, which is why we focus on 4 different few node networks to evaluate the performance of our methodology. We additionally study a futuristic scenario using a real-world fiber topology based on the core network of SURFnet, a network provider for Dutch educational and research institutions.
We implement the demand processing pipeline of our TDMA architecture using Python to evaluate the performance of several heuristics at the schedule construction step. As the performance of routing and protocol selection algorithms lie outside of the scope of this paper, we implement these stages using a shortest-path routing algorithm (edge weights corresponding to link length) and our own extension to a previously studied protocol selection algorithm known as entanglement swapping search scheme  \cite{bacinoglu2010constant}. We do not expect that the choice of these algorithms significantly affects our evaluation of our scheduling methods, though we remark that more refined methods for choosing a repeater protocol could be used (see e.g.~\cite{LukinPNAS, KennethPaper, azuma2021tools}).  Furthermore, we refrain from comparing the run-time performance of our scheduling heuristics as our implementation in Python is not representative of one that may be deployed in practice.

% GENERAL COMMENTS
% Simulation of the demand processing pipeline allows evaluating different implementations of each stage in tandem. This can be used to study new routing algorithms, protocol selection algorithms, and TDMA scheduling algorithms that suit the traffic patterns or topologies of different quantum networks. Furthermore, quantum network devices are still in experimental stages and no quantum networks capable of end-to-end entanglement production have been realized, thus using simulations can provide insight into the theoretical performance of quantum networks composed of several different network device platforms. To gain insight on the behavior of small-scale NISQ networks, we evaluate our demand processing pipeline on several small network topologies and observe scheduler performance under varied network load.

% We first discuss the assumed physical attributes of the devices in our simulated networks and briefly describe our concrete repeater protocol generation methodology. We then discuss effects of the choice of schedule slot size before comparing the results from our scheduling simulations.

\subsection{Hardware and Protocol Selection}
Our small network simulations use hardware-validated parameters of processing nodes based on NV centers in diamond~\cite{hensen2015loophole,dahlberg2019link}, since this platform is presently the only one in which at least three nodes are connected~\cite{upcomingGHZ}. In our simulations, each quantum network node has a single communication qubit and three storage qubits for storing entangled links. Link capabilities between network nodes are found by simulating entanglement establishment using software provided by the authors of \cite{dahlberg2019link} in Table~\ref{tab:phys_params}.  For our SURFnet simulation, we assume entanglement may be generated with fidelity $F=0.999$ at a rate of $1.4$ kHz between directly connected nodes. 
We choose the latency of repeater protocol operations in our simulations based on experimentally realized values: swapping operation $1$ ms, distillation $526$ $\mu$s, and move $961$ $\mu$s \cite{upcomingGHZ, kalb2017entanglement}. For simplicity in protocol selection, we here assume operations and memory are perfect (i.e. do not further reduce the quality of the entanglement generated), and that the entangled links correspond to a worst-case quantum state (see Appendix \ref{sec:app_rep_protocols}). Since this does not impact the evaluation of the scheduling algorithms: protocol selection makes sure that end-to-end fidelity requirements are met using the repeater protocol model we described earlier.  For lower fidelity, repeater protocols can meet QoS requirements using elementary link generation and entanglement swapping whereas higher fidelity requirements will include distillation.

%Protocols for lower fidelity are achieved using elementary link generation and entanglement swapping whereas distillation is present in protocols that must meet higher fidelity requirements.

\begin{table}[ht!]
	\centering
	\begin{tabular}{|l|l|}
		\hline
		Link Length             & $5$ km  \\ \hline
		\begin{tabular}[c]{@{}l@{}} Link Capabilities\cite{dahlberg2019link} \\ (Fidelity, Rate in Hz) \end{tabular}  & \begin{tabular}[c]{@{}l@{}} (0.88, 14.16), (0.83, 20.84), \\ (0.79, 27.83), (0.75, 33.98), \\ (0.7, 39.18), (0.66, 45.6), \\ (0.62, 51.26), (0.57, 57.73) \end{tabular} \\ \hline
		\# Comm. Qubits & $1$ \\ \hline
		\# Storage Qubits       & $3$ \\ \hline
	\end{tabular}
	\caption{Physical device parameters for small network simulations.}
	\label{tab:phys_params}
\end{table}

We also make the simplifying assumption that the nodes may perform any operation on different qubits in parallel. This does not hold for the NV platform, but again simplifies the protocol selection phase without impacting the study of scheduling procedures. Finally,we let distillation succeed with unit probability: this merely impacts the number of repetitions until end-to-end entanglement is achieved and while affecting the absolute achievable throughput, it does not impact the comparison of the throughput achieved by different scheduling methods as we aim to study here.

\begin{figure}[hb!]
	\centering
	\begin{subfigure}{\linewidth}
		\centering
		\caption{}
		\label{fig:net_topology}
		\includegraphics[width=0.4\linewidth]{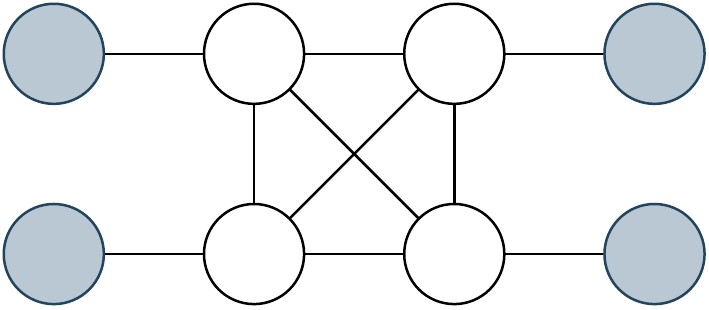}
	\end{subfigure}

	\begin{subfigure}{\linewidth}
		\caption{}
		\label{fig:surfnet_topology}
		\includegraphics[width=\linewidth]{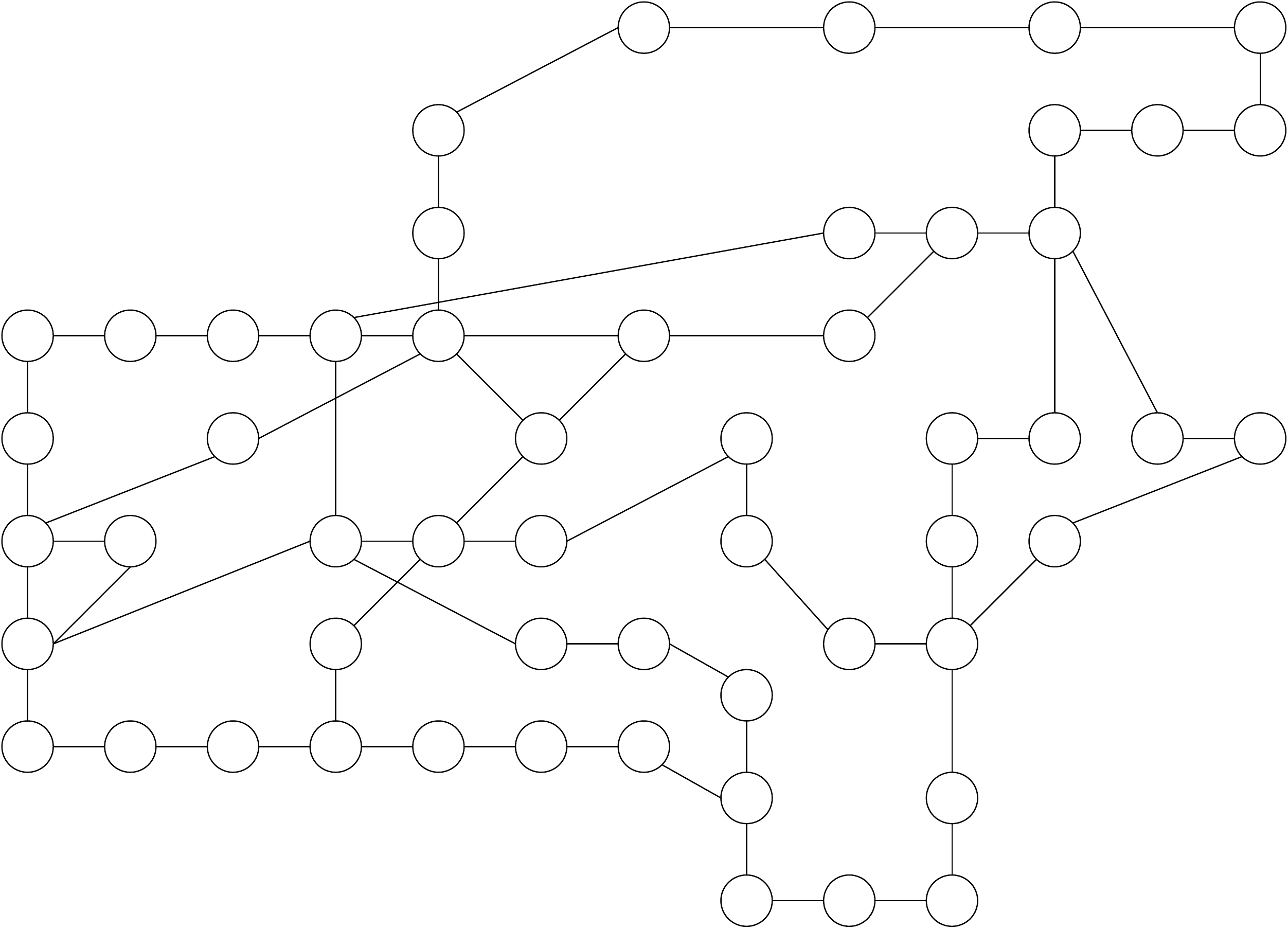}
	\end{subfigure}
	\caption{a) Network topology in scheduler evaluation. As reference, a single repeater protocol delivering entanglement of $F=0.55$ ($F=0,85$) to two end nodes (grey) via the repeaters (white) operates with latency of $170$ ($292$) ms and throughput of $5.88$ ($0.342$) $\frac{ebit}{s}$. b) SURFnet repeater topology in scheduler evaluation. Each repeater pictured (circle) is connected to an end-node that lies $5$ km away (end node not pictured).}
\end{figure}

\begin{figure*}[t!]
	\centering
	\begin{subfigure}{0.48\linewidth}
		\caption{}
		\includegraphics[width=\linewidth]{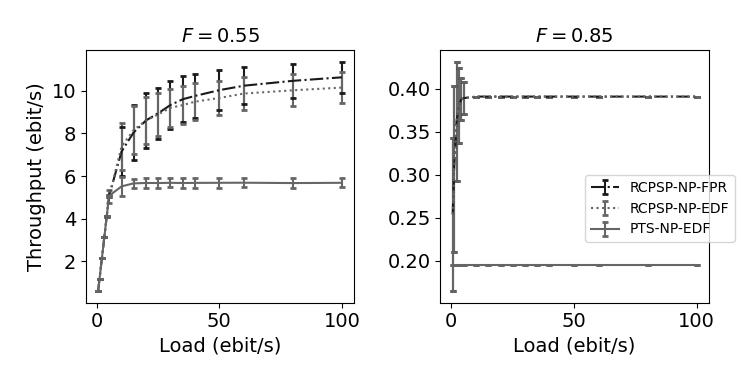}
		\label{fig:symm_throughput_v_load}
	\end{subfigure}
	\hspace{0.02\linewidth}%
	\begin{subfigure}{0.48\linewidth}
		\caption{}
		\includegraphics[width=\linewidth]{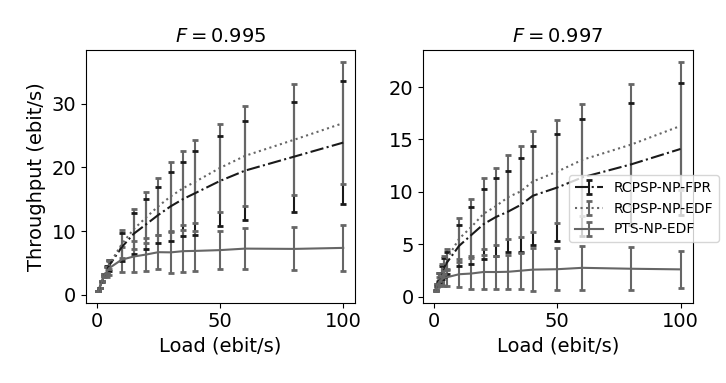}
		\label{fig:throughput_v_load_surfnet}
	\end{subfigure}
	\caption{Average network throughput vs. network load for several fidelity requirements $F$ on a) the small network in Fig. \ref{fig:net_topology} and b) the SURFnet topology in figure \ref{fig:surfnet_topology}. Data points averaged over 1000 simulations, error bars one standard deviation. RCPSP heuristics beat the throughput of the trivial schedule obtained by scheduling one single repeater protocol after the other.}
	\label{fig:throughput_v_load}
\end{figure*}

\begin{figure*}[t!]
	\centering
	\begin{subfigure}{0.48\linewidth}
		\caption{}
		\includegraphics[width=\linewidth]{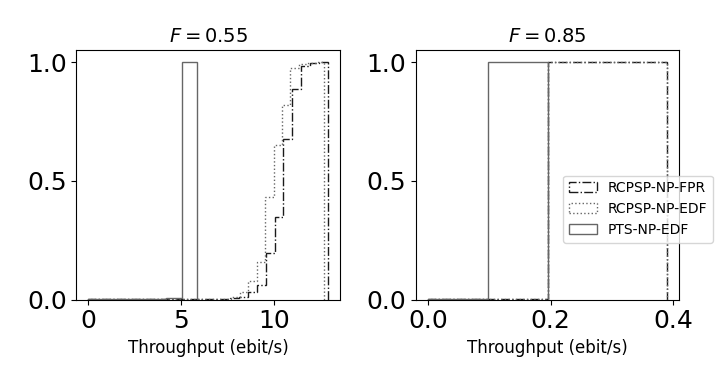}
		\label{fig:throughput_cdfs_symm}
	\end{subfigure}
	\hspace{0.02\linewidth}%
	\begin{subfigure}{0.48\linewidth}
		\caption{}
		\includegraphics[width=\linewidth]{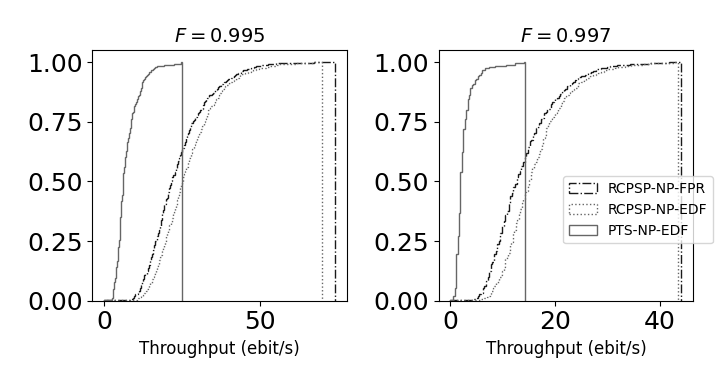}
		\label{fig:throughput_cdfs_surfnet}
	\end{subfigure}
	\caption{CDF of network throughput for a load of $100\frac{ebit}{s}$ on a) the small network of Fig. \ref{fig:net_topology} and b) the SURFnet topology. }
	\label{fig:throughput_cdfs}
\end{figure*}

\subsection{Scheduler Evaluation}
\label{sec:sched_eval}
We numerically evaluate the scaling of achieved network throughput with network load, as well as the trade-off between throughput and jitter of our implemented scheduling heuristics. Prior studies have shown that throughput decreases as the desired fidelity of an entangled link between connected devices increases \cite{dahlberg2019link}. Here we are interested in the relationship between the fidelity requirements of network demands and the achievable network throughput as well as the observed jitter in entanglement delivery. To gain additional insight on the effects of network structure on scheduler performance, we simulate the schedule construction on three additional small networks (code and data may be found online \cite{skrzypczykGithubRepo,skrzypczykDataRepo}).

\begin{figure*}[t!]
	\centering
	\begin{subfigure}{0.49\linewidth}
		\caption{}
		\label{fig:throughput_v_jitter_symm}
		\includegraphics[width=\linewidth]{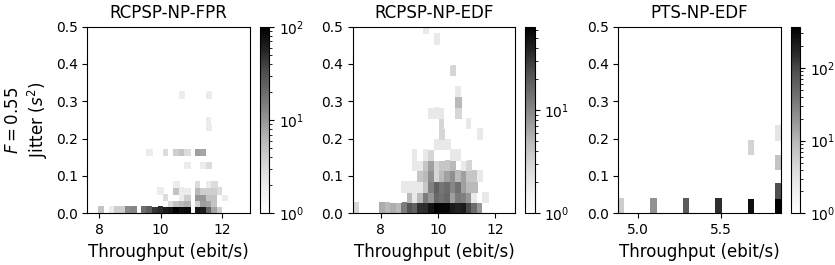}
	\end{subfigure}
	\begin{subfigure}{0.49\linewidth}
		\caption{}
		\label{fig:throughput_v_jitter_surfnet}
		\includegraphics[width=\linewidth]{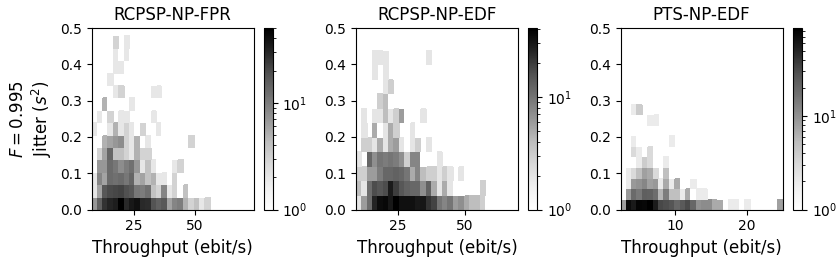}
	\end{subfigure}
	\caption{Histograms of throughput and jitter for a) the small network with fidelity requirement $F=0.55$ and b) SURFnet with a fidelity requirement of $F=0.995$ under a network load of $100 \frac{ebit}{s}$.}
	\label{fig:throughput_v_jitter}
\end{figure*}

\begin{table}[h!]
	\centering
	\begin{tabular}{|l|l|}
		\hline
		Time Slot Size          & $10$ ms \\ \hline
		Fidelity (small nets.) & $0.55, 0.6, 0.65, 0.7, 0.75, 0.8, 0.85, 0.9$ \\ \hline
		Fidelity (SURFnet) & $0.98, 0.985, 0.99, 0.995, 0.997$ \\ \hline
		Throughput ($\frac{ebit}{s}$) & \begin{tabular}[c]{@{}l@{}} 12.5, 6.25, 3.125, 1.5625, 0.78125,\\ 0.390625, 0.1953125 \end{tabular} \\ \hline
	\end{tabular}
	\caption{Slot duration and permitted QoS levels of network demands for scheduling simulations.}
	\label{tab:sim_params}
\end{table}

\subsubsection{Simulation Parameters}
Table \ref{tab:sim_params} shows the set of QoS parameters used for our simulations while Fig. \ref{fig:net_topology} shows one of the networks used for our simulations. We choose this topology due to its symmetric structure: for each path between two end nodes there exists a disjoint path and subsequently a disjoint repeater protocol connecting the remaining nodes.

We generate a batch of network demands by randomly sampling pairs of end nodes to be source and destination. Each batch of network demands has a fixed fidelity and is randomly assigned a rate from those in Table \ref{tab:sim_params}. We restrict rates to a fixed set to reduce the length of the schedule. We use a slot size of $10$ ms, for which timing synchronization to slot boundaries across network nodes may be reasonably achieved, resulting in a schedule with a maximum length of $512$ slots (equivalent to $5.12$ s).  After generating a batch of demands, we compute the path for each demand and generate repeater protocols using our modified ESSS algorithm. The set of demands and their repeater protocols are then fed to our scheduling heuristics.

\subsubsection{Scaling with Network Load}
We evaluate how the network throughput of the scheduling heuristics scales with the cumulative throughput requirements of all network demands (henceforth referred to as \emph{network load}). Evaluating schedulers on this basis is important as achieving higher network throughput shows how well each scheduler extracts parallelism from the network. In these simulations, we generate batches of demands for several network loads and apply our scheduling heuristics. Increasing the number of demands in the system provides schedulers an opportunity to squeeze additional throughput from the network if some portions of the network are not fully utilized.

Fig. \ref{fig:throughput_v_load} shows the average network throughput achieved by each scheduler for several choices of end-to-end fidelity. Both RCPSP scheduling heuristics achieve higher network throughput than our periodic task scheduling heuristic under high load and show comparable performance for low network loads. This confirms our expectations of achieving higher network performance when using the RCPSP scheduling method. We observe that the achievable network throughput decreases as fidelity requirements are increased which agrees with the results in \cite{dahlberg2019link}. These results additionally suggest that the central controller may choose to use RCPSP heuristics for higher network load while using periodic task scheduling heuristics for low network load in order to reduce scheduling overhead while maintaining network performance. The large variance observed in the average throughput in our SURFnet simulations is the result of network demands that highly congest a common repeater, resulting in lower network throughput.

% As fidelity requirements increase, the difference in network throughput of RCPSP scheduling heuristics decreases. For fidelity requirements of 0.85 the RCPSP scheduling heuristics achieve the same average throughput once network load surpasses 10 $\frac{ebit}{s}$.

From our simulations we also learn how well our novel RCPSP-NP-FPR heuristic compares to the PTS-NP-EDF and RCPSP-NP-EDF heuristics.  Fig. \ref{fig:throughput_cdfs} shows the CDF of achieved network throughput when the network load is fixed to be $100 \frac{ebit}{s}$. Here, we see that the throughput distribution of RCPSP-NP-FPR is higher than that for PTS-NP-EDF and comparable to that of the RCPSP-NP-EDF heuristic.  This shows that our heuristic can maintain high performance while reducing computational complexity as compared to RCPSP-NP-EDF. These results are further supported by our simulations on other small networks \cite{skrzypczykDataRepo}.

%XXX NOT SURE HOW TO PHRASE THIS: The repeater protocol for delivering entanglement of $F=0.55$ ($F=0,85$) to end nodes on the symmetric network operates with a latency of $170$ ($292$) ms corresponding to a throughput of $5.88$ ($0.342$) $\frac{ebit}{s}$. Since two such protocols can operate on disjoint paths in the symmetric topology, we may estimate a lower bound on the maximum throughput for $F=0.55$ to be $11.76$ ($0.684$) $\frac{ebit}{s}$. In Fig. \ref{fig:throughput_v_load} we see that for $F=0.55$ the RCPSP scheduling heuristics can meet and even exceed this quantity as repeater protocol executions can be scheduled to overlap on non-disjoint paths. As for $F=0.85$, we see that RCPSP heuristics reach a rate of approximately $0.38$ $\frac{ebit}{s}$ for higher network loads. Since the latency of the repeater protocol for $F=0.85$ has a latency of $292$ ms it can only be used to satisfy rate requirements of $0.19$ (Table \ref{tab:sim_params}) resulting in a throughput of $0.38$ when repeater protocols are executed in parallel. XXX
 
%The results from our four networks (see Appendix \ref{sec:app_results}) show that the RCPSP-CEDF scheduling heuristic generally outperforms the remaining heuristics in terms of network throughput for lower fidelity requirements while the distinction is less pronounced for higher fidelity requirements. As fidelity requirements increase, we observe that our RCPSP-FPR heuristic can perform comparably or even better than RCPSP-CEDF and RCPSP-NP-EDF while reducing the runtime complexity.

\subsubsection{Throughput/Jitter Trade-Off}
In addition to network throughput, we extract the jitter in entanglement delivery. High amounts of jitter result in irregular entanglement delivery and may affect applications that fall into the \emph{Create and Keep} use case \cite{dahlberg2019link} negatively.  While the evaluated scheduling heuristics are not targeted at minimizing jitter, it is instructive to characterize the trade-off to the achieved throughput. We remark that any demand $D_i=(A, B, F_{min}, R_{min}, J_{max})$ with $J_{max} \leq \frac{1}{(R_{min})^2}$ is guaranteed to be satisfied using either the periodic task scheduling or RCPSP methods we propose.

Fig. \ref{fig:throughput_v_jitter} shows the distribution of throughput/jitter pairs for our scheduling heuristics when network load is $100 \frac{ebit}{s}$. From our results we learn that our periodic task scheduling heuristic observe smaller amounts of jitter compared to the RCPSP heuristics in our small network simulations, suggesting that achieving high network throughput comes at a cost of increased variance in inter-delivery times of entanglement.  However, we see that for the SURFnet topology, where the path length between pairs of end-nodes varies, that the periodic task scheduling and RCPSP methods show comparable levels of jitter. This is further supported by our additional simulations of small networks.

\subsection{Summary}
Our simulation results show that our architecture can be used to accommodate different levels of network performance by controlling the choice of scheduling strategy. We show that flexibility in the choice of scheduling heuristic allows network operators to control trade-offs between scheduling overhead, achievable network throughput, and observed jitter in entanglement delivery.
%For low fidelity requirements, RCPSP-based heuristics are more suitable for achieving higher network throughput whereas periodic task-scheduling heuristics are more suitable for achieving lower jitter. In this realm, we also show that our novel RCPSP-FPR heuristic can be used to reach comparable levels of performance to RCPSP-NP-EDF while reducing scheduling complexity. At higher fidelity requirements, the distinction between these scheduling methods becomes less pronounced and the use of periodic task scheduling heuristics can provide the same performance as RCPSP-based heuristics with a significantly lower overhead.
When end-to-end fidelity requirements can be met with low-latency repeater protocols, RCPSP-based heuristics are more suitable for achieving higher network throughput whereas periodic task-scheduling heuristics are more suitable for achieving lower jitter. In this realm, we also show that our novel RCPSP-FPR heuristic can be used to reach comparable levels of performance to RCPSP-NP-EDF while reducing scheduling complexity. However, when repeater protocols have high latency, due to spending more time producing entangled links, the distinction between these scheduling methods with respect to network throughput and jitter becomes less pronounced.

Our results on other network topologies \cite{skrzypczykDataRepo} further support these observations and additionally provide insight into the effects of network topology on network throughput. We find that network structures with non-uniform path lengths between end nodes result in larger amounts of variance in the achievable network throughput and that periodic task scheduling heuristics achieve network throughput comparable to RCPSP heuristics on a star topology.

In order to abstract away from the details of the physical and link layer protocols, we remind that for simplicity we presented our work using a time granularity such that sufficiently many time slots are allocated for entanglement generation (or other probabilistic operations) to succeed with high probability. We remark that this is clearly wasteful when producing many entangled pairs, and a more fine grained allocation of time slots can be beneficial in practise, depending on user demand, demands to intersperse local operations at the end nodes, the reactiveness of the link and physical layer implementation - to name but a few of the factors that may inform a choice of time granularity. Our TDMA architecture and general methods can however be used with many choices of time granularity, and we leave a choice of time granularity for specific use cases for future research.

In this work we have presented the first end-to-end design of a centralized quantum network architecture that supports varying levels of QoS requirements among multiple pairs of users.  Our architecture may be used to scale near-term networks and may additionally be used to manage smaller networks to form a larger, cluster-based network of quantum devices.  Our results may additionally be used as a baseline to evaluate the performance of alternative architecture designs that may target a more distributed approach to network coordination.

This work does not raise any ethical issues.

% \section{Open Questions}
%\label{sec:conc}
%Several avenues for future research stem from the work presented. First, the presented demand processing pipeline isolates the routing, repeater protocol selection, and TDMA schedule construction stages. This modularity enables evaluation of alternative algorithms and allows flexibility in network design to suit expected traffic patterns. Second, performing routing, protocol selection, and scheduling jointly can potentially improve network performance as choices of routes and repeater protocols can be computed based on previous scheduling decisions. Third, centralized network management will not scale well with larger networks, thus the design of distributed or clustered TDMA scheduling protocols are of particular interest.
%
%\section{Ethical Considerations}
%This work does not raise any ethical issues.

% \section*{Acknowledgments}

\FloatBarrier

\appendix

\section{Appendix}
In this Appendix we briefly describe our method for generating repeater protocols as input to our simulations as well as extensions to our scheduling methods which allow for handling jitter QoS requirements.  Additional information, code, and data analysis may be found online at \cite{skrzypczykDataRepo}.

\begin{figure*}[h!]
	\centering
	\includegraphics[width=0.6\linewidth]{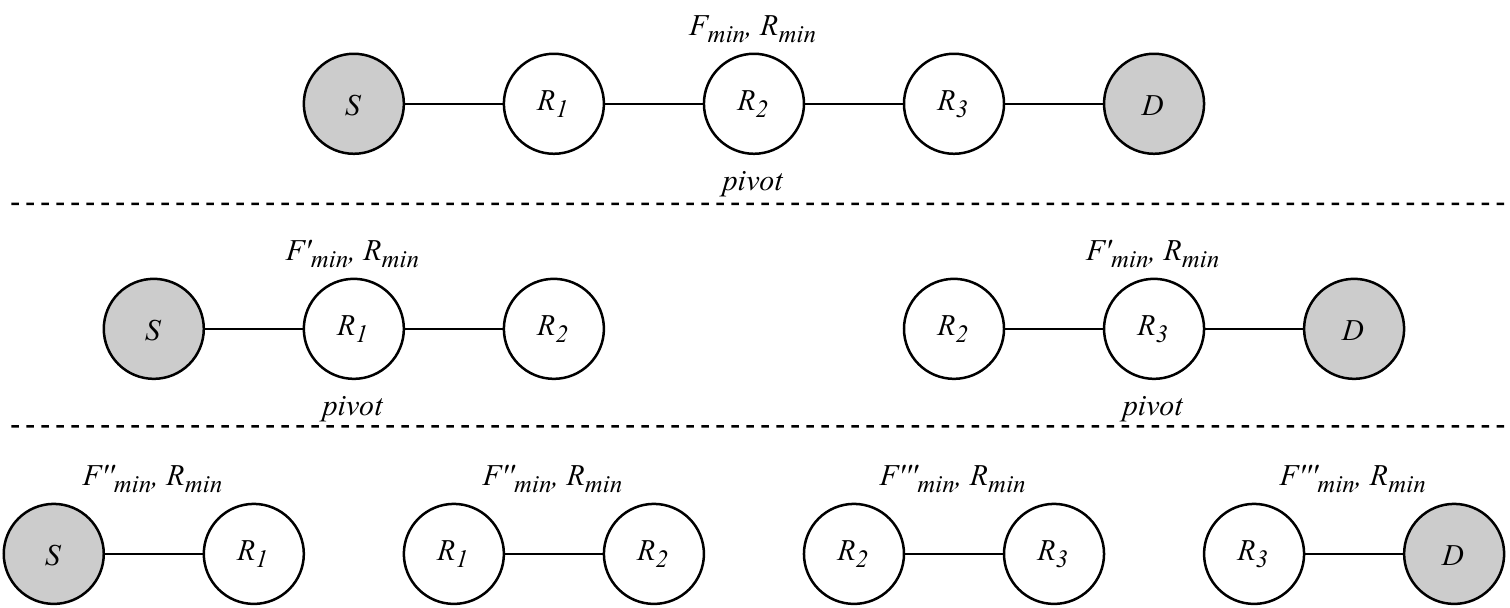}
	\caption{Visual depiction of the operation of ESSS. The algorithm begins with the full repeater chain (top). A pivot is then chosen and the chain is split in two (middle). New pivots are chosen at each level until elementary links remain (bottom).}
	\label{fig:esss_example}
\end{figure*}

\subsection{Repeater Protocol Generation}
\label{sec:app_rep_protocols}
In our simulations we have assumed that the state of entangled links are of the Werner form \cite{werner1989quantum} which corresponds to a "worst-case" quantum state that is a probabilistic mixture of the perfectly entangled state $|\Phi^+\rangle$ and a separable (not entangled) state $\mathbb{I}$

\begin{equation}
\rho_{werner} = \frac{1-F}{3}\mathbb{I}_2 + \frac{4F-1}{3}|\Phi^+\rangle\langle\Phi^+|
\end{equation}

The fidelity $F_S$ of a link produced from entanglement swapping with two entangled links of the Werner form with fidelity $F_1$ and $F_2$ can be computed as

\begin{equation}
F_S = F_1 F_2 + \frac{(1-F_1)(1-F_2)}{3},
\label{eq:werner_swap_fidelity}
\end{equation}

\noindent while the fidelity $F_D$ of a link produced using two-to-one entanglement distillation can be computed as

\begin{equation}
F_D = \frac{F_1 F_2 + \frac{(1-F_1)}{3}\frac{(1-F_2)}{3}}{F_1 F_2 + F_1 \frac{(1-F_2)}{3} + F_2 \frac{(1-F_1)}{3} + 5\frac{(1-F_1)}{3}\frac{(1-F_2)}{3}}.
\label{eq:werner_distill_fidelity}
\end{equation}

To generate repeater protocols for our simulations, we extend the entanglement swapping scheme search (ESSS) algorithm \cite{bacinoglu2010constant} to consider both entanglement swapping and entanglement distillation at the pivot node level. ESSS may be understood in the following way: given a path of quantum network nodes starting at a source $S$ and ending at a destination $D$ along with a minimum fidelity requirement $F_{min}$ and rate requirement $R_{min}$, a "pivot" node $P$ is chosen internal to the path and repeater protocols are recursively found on the two induced subpaths on either side of the pivot. A repeater protocol for delivering entanglement to $S$ and $D$ may be constructed by performing an entanglement swap at the pivot node $P$ using the links produced by a repeater protocol on either side of $P$. A repeater protocol is then found recursively on each of the subpaths in a similar fashion. An example of the ESSS algorithm may be seen in Fig. \ref{fig:esss_example}.

Since performing an entanglement swap at the pivot reduces the fidelity of the produced link, at each level of ESSS we update the required fidelity $F'_{min}$ of each repeater protocol on either side of the pivot.  Using our Werner state assumption and \ref{eq:werner_swap_fidelity}, we choose $F'_{min}$ such that

\begin{equation}
F_{min} = F'^2_{min} + \frac{(1-F'_{min})^2}{3}.
\end{equation}

Our search assumes two-to-one entanglement distillation is performed and considers two styles of entanglement distillation known as \emph{entanglement pumping} and \emph{nested entanglement pumping} \cite{briegel1998quantum, dur1999quantum, dur2003entanglement}. Standard entanglement pumping attempts to produce a link of high fidelity by repeatedly performing two-to-one entanglement distillation using a long-running link and new entanglement as it is produced. This style of entanglement distillation has minimal resource requirements as only two links are used at any time.  Nested entanglement pumping, on the other hand, performs two-to-one entanglement distillation using pairs of links with lower fidelity before distilling links with higher fidelity together as shown in. Depending on the desired depth of nesting, the number of links, and consequently the minimum resource requirements, increases. The trade-off between these two methods of entanglement distillation is that nested entanglement pumping can be used to reach near-perfect ($F \approx 1$) fidelity while non-nested entanglement pumping reaches an asymptotic limit on achievable fidelity depending on the fidelity of the fidelity of new entanglement.

The general framework of the ESSS provides a method for searching for quantum repeater protocols but does not provide a means for mapping them to qubits held by the network devices. Some network devices may be restricted to producing entangled links serially while others can produce multiple links concurrently. Characterizing the minimum latency, and hence the rate, of quantum repeater protocol execution thus requires a temporal mapping of operations in the protocol to qubits in network devices.  This allows one to determine the amount of time needed to execute the quantum repeater protocol and if the protocol meets rate requirements.

Since the set of operations $A$ in a quantum repeater protocol have a consumer/producer relationship and the dependency relationship on operations takes the form of a tree structure, we determine the resource mapping $Q$ and relative offset mapping $M$ for the repeater protocol $P(A,I,M,Q)$ by performing a post-order traversal of the protocol operation tree.  When a source (link establishment operation) node $a$ is reached when traversing the tree, the operation is mapped to communication qubits and storage qubits held by the network nodes described by $a_V$. In our model (Section \ref{sec:sched}), we assume that any communication qubit may move a link into any storage qubit held by the same node. We thus always assign a vacant communication qubit and a vacant storage qubit to the operation so that a newly established link may immediately be stored and free the communication qubit for subsequent link establishment. When no storage qubits are available, the entangled link remains in the communication qubit until it is freed by an entanglement swap or entanglement distillation that consumes the held link. $Q(a)$ for a link establishment operation $a$ is then the set of communication qubits and storage qubits chosen for an operation $a$. The assigned qubits are propagated through the protocol DAG to entanglement swaps and entanglement distillation operations that consume the qubits. This ensures that the entangled links in the qubits are correctly paired with one another for entanglement swapping and entanglement distillation.

To determine $M$, we produce a "mini" schedule for the set of operations $A$. We track the vacancy/occupation of communication qubits and storage qubits in time as they are used by quantum repeater protocol operations. Each operation $a \in A$ for a quantum repeater protocol is allocated an appropriate amount of time to succeed with high probability. For link establishment, we set this time to the expected latency of entanglement establishment between the two connected network nodes $u,v \in a_V$ for the given fidelity $a_F$. Our quantum repeater protocols aim to produce a single entangled link per operation, thus operations for link establishment are allocated enough slots to produce one entangled link. For entanglement swapping and entanglement distillation operations, the allocated amount of time depends on device capabilities in the network. Operations for entanglement swapping and entanglement distillation need only be allocated enough slots in the schedule to execute the operation once.

To reduce protocol latency and preserve link fidelity, operations are scheduled in two passes. The first pass determines an as-soon-as-possible (ASAP) scheduling of the protocol operations. This determines the latency of the protocol and finds an initial scheduling. The schedule is then processed again to push link establishment as-late-as-possible (ALAP) while keeping all entanglement swap and entanglement distillation operations in place. Doing so reduces that amount of time a link may need to be stored before being consumed by one of these operations.

\subsection{Scheduling Extensions for Handling Jitter}
\label{sec:app_sched_ext}
The scheduling methods we have considered in Section \ref{sec:sched} are designed to satisfy throughput requirements on entanglement delivery while jitter requirements are met on a best-effort basis.  Both of our proposed methods may be extended to additionally handle jitter requirements.

The periodic task scheduling method may be extended to handle jitter constraints by introducing relative timing constraints between the instances of each periodic task \cite{cheng1995allocation}.  In this case, each periodic task $\tau_i=(\Phi_i, C_i, T_i)$ is associated with timing constraints $\lambda_i,\eta_i$ that require each pair of consecutive instances $\tau_{i,j}$, $\tau_{i,j+1}$ with start times $s_{i,j}$, $s_{i,j+1}$ (in slots) to satisfy $s_{j,j+1} \leq s_{i,j} + T_i + \eta_i$ and $s_{i,j+1} \ge s_{i,j} + T_i - \lambda_i$.

Using relative timing constraints, one may specify $\lambda_i=\eta_i=\lfloor\frac{\sqrt{J_{i,max}}}{t_{slot}}\rfloor$ for each periodic task $\tau_i$ in order to satisfy jitter requirements due to the fact that the observed jitter $J_i$ for a periodic task is:

\begin{align}
J_i &= \frac{\sum_{j=1}^{|\mathcal{S}(D_i, P_i)|} (s_{j+1}t_{slot}-s_jt_{slot} - \bar{s}_it_{slot})^2}{|\mathcal{S}(D_i, P_i)|}\\
&= \frac{\sum_{j=1}^{|\mathcal{S}(D_i, P_i)|} (s_{j+1}t_{slot}-s_jt_{slot} - T_it_{slot})^2}{|\mathcal{S}(D_i, P_i)|}\\
&\le \frac{|\mathcal{S}(D_i, P_i)|(T_it_{slot} + \lambda_it_{slot} - T_it_{slot})^2}{|\mathcal{S}(D_i, P_i)|} \\ 
&= (\lambda_it_{slot})^2 \\
&\le J_{i,max}
\end{align}

\noindent where we have assumed without loss of generality that $\eta_i \le \lambda_i$.

For RCPSP, we apply a similar principle by specifying additional minimal/maximal time lags between the dummy start activities of consecutive instances of the activity-on-node network for each repeater protocol.  Specifically, to satisfy a maximum jitter requirement of $J_{i,max}$ we set $d_{s_{i,j}s_{i,j+1}}=T_i-\lfloor\frac{\sqrt{J_{i,max}}}{t_{slot}}\rfloor$ and $\bar{d}_{s_{i,j}s_{i,j+1}}=T_i+\lfloor\frac{\sqrt{J_{i,max}}}{t_{slot}}\rfloor$ between the dummy start activities $j_{s_{i,j}}$ and $j_{s_{i,j+1}}$ for the $j$th and $(j+1)$th instance of the activity-on-node network for repeater protocol $P_i$.  Adding these timing constraints satisfies the jitter requirements due to similar reasoning as in the case for periodic task scheduling.

\bibliographystyle{IEEEtran}
\bibliography{references}

\end{document}